\documentclass[twocolumn]{aastex62}
\usepackage{lineno}
\UseRawInputEncoding
\hypersetup{linkcolor=red,citecolor=blue,filecolor=cyan,urlcolor=magenta}

\usepackage{xcolor}
\usepackage{amsmath}
\usepackage{graphicx}   
\usepackage{bm}         
\usepackage{natbib}
\usepackage{etoolbox}
\usepackage{rotating}
\usepackage{array}
\usepackage{booktabs}
\usepackage{amssymb}
 
\newcommand{\msun}{\mbox{\,$M_\odot$}}

\newcommand{\msunyr}{\mbox{\,$M_\odot$ yr$^{-1}$}}
\newcommand{\kms}{\mbox{\,km\,s$^{-1}$}}

\newcommand{\ee}[1]{\mbox{${} \times 10^{#1}$}}

\newcommand{\methylformate}{\mbox{CH$_{3}$OCHO}}

\newcommand{\methanol}{\mbox{CH$_{3}$OH}}
\newcommand{\tmethanol}{\mbox{$^{13}$CH$_{3}$OH}}

\newcommand{\dmethanol}{\mbox{CH$_{2}$DOH}}

\newcommand{\dimethylether}{\mbox{CH$_{3}$OCH$_{3}$}}

\newcommand{\acetaldehyde}{\mbox{CH$_{3}$CHO}}
\newcommand{\ethylcyanide}{\mbox{CH$_{3}$CH$_{2}$CN}}

\newcommand{\sotwo}{\mbox{SO$_{2}$}}

\newcommand{\formamide}{\mbox{NH$_{2}$CHO}}

\newcommand{\htwo}{\mbox{$\rm H_2$}}
\newcommand{\htwoco}{\mbox{$\rm H_2CO$}}
\newcommand{\thirteenmethanol}{\mbox{$\rm ^{13}CH_3OH$}}

\newcommand{\ceighteeno}{\mbox{$\rm C^{18}O$}}
\newcommand{\thirteenco}{\mbox{$\rm ^{13}CO$}}

\newcommand{\so}{\mbox{$\rm SO$}}

\newcommand{\gcm}{\mbox{g cm$^{-2}$}} 
 
\newcommand{\mjypbm}{\mbox{mJy\,beam$^{-1}$}}
\newcommand{\mujypbm}{\mbox{$\mu$Jy\,beam$^{-1}$}}

\received{}
\revised{}
\accepted{}

\shorttitle{Sample article}
\shortauthors{Cheng et al.}

\begin{document}

\title{Disks and Outflows in the Intermediate-mass Star Forming Region NGC~2071~IR}

\correspondingauthor{Yu Cheng}
\email{ycheng.astro@gmail.com}

\author[0000-0002-8691-4588]{Yu Cheng}
\affiliation{National Astronomical Observatory of Japan, 2-21-1 Osawa, Mitaka, Tokyo, 181-8588, Japan}
\affiliation{National Radio Astronomy Observatory, 520 Edgemont Rd., Charlottesville, VA 22903, USA}

\author[0000-0002-6195-0152]{John J. Tobin}
\affiliation{National Radio Astronomy Observatory, 520 Edgemont Rd., Charlottesville, VA 22903, USA}

\author[0000-0001-8227-2816]{Yao-Lun Yang}
\affiliation{RIKEN Cluster for Pioneering Research, Wako-shi, Saitama, 351-0106, Japan}
\affiliation{Department of Astronomy, University of Virginia, Charlottesville, VA 22903, USA}

\author{Merel L. R. van 't Hoff}
\affiliation{University of Michigan, Department of Astronomy, 1085 S. University, Ann Arbor, MI 48109, USA}

\author{Sarah I. Sadavoy}
\affiliation{Department  for  Physics,  Engineering  Physics  and  Astrophysics, Queen’s University, Kingston, ON, K7L 3N6, Canada}

\author[0000-0002-6737-5267]{Mayra Osorio}
\affiliation{Instituto de Astrof\'{\i}sica de Andaluc\'{\i}a, CSIC, Glorieta de la Astronom\'{\i}a s/n, E-18008 Granada, Spain}

\author[0000-0001-9112-6474]{Ana Karla D{\'i}az-Rodr{\'i}guez }
\affiliation{UK ALMA Regional Centre Node, Jodrell Bank Centre for Astrophysics, Department of Physics and Astronomy, The University of Manchester, Oxford Road, Manchester M13 9PL, UK.}

\author[0000-0002-7506-5429]{Guillem Anglada}
\affiliation{Instituto de Astrof\'{\i}sica de Andaluc\'{\i}a, CSIC, Glorieta de la Astronom\'{\i}a s/n, E-18008 Granada, Spain}

\author[0000-0003-3682-854X]{Nicole Karnath}
\affiliation{SOFIA Science Center, USRA, NASA Ames Research Center, Moffett Field, CA 94035, USA}

\author{Patrick D. Sheehan}
\affiliation{Northwestern University, Evanston, IL}

\author{Zhi-Yun Li}
\affiliation{Department of Astronomy, University of Virginia, Charlottesville, VA 22903}

\author{Nickalas Reynolds}
\affiliation{Homer L. Dodge Department of Physics and Astronomy, University of Oklahoma, 440 W. Brooks Street, Norman, OK 73019, USA}

\author{Nadia M. Murillo}
\affiliation{Star and Planet Formation Laboratory, RIKEN Cluster for Pioneering Research, Wako, Saitama 351-0198, Japan}

\author{Yichen Zhang}
\affiliation{Star and Planet Formation Laboratory, RIKEN Cluster for Pioneering Research, Wako, Saitama 351-0198, Japan}

\author{S. Thomas Megeath}
\affiliation{Department of Physics and Astronomy, University of Toledo, Toledo, OH 43560}

\author{Łukasz Tychoniec}
\affiliation{European Southern Observatory, Karl-Schwarzschild-Strasse 2, 85748 Garching bei M\"unchen, Germany}

\begin{abstract}

We present ALMA band 6/7 (1.3 mm/0.87 mm) and VLA Ka band (9 mm) observations toward NGC 2071 IR, an intermediate-mass star forming region. We characterize the continuum and associated molecular line emission towards the most luminous protostars, i.e., IRS1 and IRS3, on $\sim$ 100 au (0\farcs{2}) scales. IRS1 is partly resolved in millimeter and centimeter continuum, which shows a potential disk. IRS3 has a well resolved disk appearance in millimeter continuum and is further resolved into a close binary system separated by $\sim$40 au at 9 mm. Both sources exhibit clear velocity gradients across their disk major axes in multiple spectral lines including \ceighteeno, \htwoco, \so, \sotwo, and complex organic molecules like \methanol, \thirteenmethanol{} and $\rm CH_3OCHO$. We use an analytic method to fit the Keplerian rotation of the disks, and give constraints on physical parameters with a MCMC routine. The IRS3 binary system is estimated to have a total mass of 1.4--1.5 \msun. IRS1 has a central mass of 3--5 \msun{} based on both kinematic modeling and its spectral energy distribution, assuming that it is dominated by a single protostar. For both IRS1 and IRS3, the inferred ejection directions from different tracers, including radio jet, water maser, molecular outflow, and \htwo{} emission, are not always consistent, and for IRS1, these can be misaligned by $\sim$50\arcdeg. IRS3 is better explained by a single precessing jet. A similar mechanism may be present in IRS1 as well but an unresolved multiple system in IRS1 is also possible.

\end{abstract}

\keywords{}

\section{Introduction}\label{sec:intro}
Intermediate mass protostars are observationally defined as young stellar objects (YSOs) that have luminosities between $\sim$ 50 and 2000 $L_{\odot}$ and will eventually reach final masses of 2--8 $M_{\odot}$ \citep{Beltran15}. Intermediate mass protostars constitute the link between low- and high-mass protostars, and hence provide a natural laboratory to test star formation theories that unify the two mass regimes. Unlike their low mass counterparts, intermediate mass stars produce significantly more UV photons and form in more densely clustered environments \citep[e.g.,][]{Fuente07}. In observational terms, intermediate mass star-forming regions are on average closer and less extincted than high mass ones, making it easier to trace the primordial configuration of the molecular cloud and to study the earliest stages of star formation.

NGC~2071~IR is an intermediate mass star-forming region located in the Orion B molecular cloud, approximately 4\arcmin{} north of the optical reflection nebula NGC~2071. The distance of NGC~2071~IR is about 430.4~pc as estimated in \citet{Tobin20}, which is based on Gaia DR2 data for a sample of relatively evolved young stars in Orion. This region is characterized by an energetic bipolar outflow, which is oriented in the NE-SW direction and extends $\sim$15\arcmin{} in length and $\sim$120 \kms{} in velocity. The outflow has been extensively characterized in CO \citep[e.g.,][]{Bally82,Scoville86,Stojimirovic08} and \htwo{} 2.12 $\mu$m emission \citep{Eisloffel00,Walther19}. At the center of the outflow is an infrared cluster with a diameter of $\sim$30\arcsec, which has a total luminosity of 520 L$_{\odot}$ \citep{Butner90}, and harbors $\sim$10 near-IR sources \citep{Persson81,Walther93,Walther19}. Most of the near-IR sources are identified as YSOs \citep{Skinner09}.

Millimeter and centimeter continuum emission has been detected toward some of the IR sources \citep{Snell86,Torrelles98,Trinidad09,VanKempen12,Carrasco12}. Among these sources IRS1 and IRS3 are of particular interest as they are the dominant mid/far-IR luminosity contributors and also presumed driving sources of the large scale outflow \citep[e.g.,][]{Torrelles98,Eisloffel00}. 
Both IRS1 and IRS3 are resolved into three components in 1.3 cm continuum emission, with the outer components interpreted as ionized gas being ejected by the central objects \citep{Trinidad09}. \citet{Carrasco12} found a variation of the elongation direction of IRS1 at 3.6 cm over 4 years, possibly indicating unobserved multiplicity inside IRS1.
In both sources, the water maser emission appears to trace parts of a rotating protostellar disk and a collimated outflow \citep{Torrelles98,Seth02, Trinidad09}. Based on the spatial-velocity distribution of masers that traces protostellar disks, \citet{Trinidad09} estimated the central mass of IRS1 and IRS3 to be $\sim$5 and $\sim$1 \msun, respectively. 

Building on these previous studies, we have conducted  Atacama Large Millimeter/submillimeter Array (ALMA) and Karl G. Jansky VLA observations at 0.87, 1.3, and 9~mm, detecting and resolving the dust and free-free emission from the protostars within the NGC~2071~IR region. Furthermore, the molecular line emission contained within our ALMA bandpass enables us to further characterize the physical conditions of the protostars in the region, and in particular, to give more stringent constraints on the dynamical masses of IRS1 and IRS3. This paper is structured as follows: the observations and results are presented in \autoref{sec:obs} and \autoref{sec:res}, respectively. We perform a spectral energy distribution (SED) analysis in \autoref{sec:sed}, and kinematic modeling of the protostellar disks in \autoref{sec:kine}. The results are further discussed in \autoref{sec:diss}, and we present our conclusions in \autoref{sec:con}.

\startlongtable
\begin{deluxetable*}{cccccc}
\tabletypesize{\scriptsize}
\renewcommand{\arraystretch}{1.0}
\tablecaption{Information on the spectral lines in our ALMA band 6 observations\label{table:line_info}}
\tablehead{
\colhead{Transition} & \colhead{Frequency}  & \colhead{$E_{\rm up}/k$} & \colhead{Beam size} & \colhead{Velocity reso.} & \colhead{RMS} \\
\colhead{} & \colhead{GHz}  & \colhead{K} & \colhead{\arcsec $\times$ \arcsec} & \colhead{\kms} & \colhead{\mjypbm{} per channel}
}
\startdata
$\rm C^{18}O~ 2-1$                   &  219.560358 & 15.8    & 0.29$\times$0.26 & 0.05 & 5.0\\
$\rm ^{13}CO~ 2-1$                   &  220.398684 & 15.9    & 0.29$\times$0.26 & 0.05 & 8.0\\
$\rm ^{12}CO~ 2-1$          &  230.538000 & 16.6    & 0.28$\times$0.24 & 1.00 & 1.6\\
$\rm H_2CO ~ 3_{0,3}-2_{0,2}$        &  218.222195 & 21.0     & 0.29$\times$0.26 & 0.20 & 2.6   \\
$\rm H_2CO ~ 3_{2,2}-2_{2,1}$        &  218.475632 & 68.1     & 0.29$\times$0.26 & 0.20 & 2.6   \\
$\rm H_2CO ~ 3_{2,1}-2_{2,0}$        &  218.760066 & 68.1     & 0.29$\times$0.26 & 0.20 & 2.6   \\
SO $\rm 6_5-5_4$                     &  219.949442 & 35.0      & 0.29$\times$0.26 & 0.20 & 3.5   \\
$\rm ^{13}CH_3OH ~ 5_{1,5}-4_{1,4}$  &  234.011580 & 48.3     &  0.26$\times$0.23 & 1.25 & 1.2 \\
$\rm CH_3OCHO ~18_{4,14}-17_{4,13}$  &  233.777515 & 114.4    &  0.26$\times$0.23 & 1.25 & 1.2\\
$\rm CH_3OH ~ 18_{3,16}-17_{4,13}$   &  232.783446 & 446.5   &  0.26$\times$0.23 & 1.25 & 1.0\\
$\rm CH_3OH ~ 10_{3,7}-11_{2,9}$     &  232.945797 & 190.4   &  0.26$\times$0.23 & 1.25 & 1.0\\
$\rm CH_3OH ~ 18_{3,15}-17_{4,14}$   &  233.795666 & 446.6   &  0.26$\times$0.23 & 1.25 & 1.0\\
$\rm SO_2~ 28_{3,25}-28_{2,26}$      &  234.187057 & 403.3   &  0.26$\times$0.23 & 1.25 & 1.2\\ 
$\rm SO_2~16_{6,10}-17_{5,13}$       &  234.421588 & 213.3   &  0.26$\times$0.23 & 1.25 & 1.2\\
\enddata
\end{deluxetable*}

\section{Observations}\label{sec:obs}
The ALMA band~7 and VLA Ka observations presented here are part of the VLA/ALMA Nascent Disk and Multiplicity (VANDAM) Survey of the Orion molecular clouds. Observations were conducted toward 328 protostars (148 for the VLA) in the Orion molecular clouds, all at $\sim$0\farcs{1} resolution. The full survey results are presented in \citet{Tobin20}. The sample of 328 protostars is derived from the HOPS sample \citep{Furlan16}, observing the bona fide protostars from Class 0 to Flat Spectrum.

\subsection{ALMA band 7 and VLA observations}
The detailed information of ALMA band~7 (0.87~mm) and VLA Ka (9~mm) observations can be found in \citet{Tobin20}. In this work we mainly utilize the continuum images. The beam sizes are 0\farcs{13} $\times$ 0\farcs{10} (56~au $\times$ 43~au) for 0.87~mm and 0\farcs{09} $\times$0\farcs{06} (39~au $\times$ 26~au) for 9~mm, respectively. The maxium recoverable scales are about 1\farcs{2} for 0.87~mm, and 1\farcs{6} for 9~mm. The 0.87~mm map has a rms noise of 0.55~\mjypbm{}, and the 9~mm continuum map has a rms noise of 12~\mujypbm.

\subsection{ALMA band 6 Observations}

NGC~2071~IR was observed with ALMA at 1.3~mm in six executions from Oct 2 to Nov 23 in 2018. The observations were conducted with 42 -- 49 operating antennas and covered baselines from 15~m to 2500~m.  
The correlator was configured with the first baseband split into two 58.6 MHz spectral windows with 1920 channels each (0.041~\kms{} velocity resolution) and centered on $^{13}$CO 2--1 and C$^{18}$O 2--1, respectively. The second baseband was split into four 58.6 MHz spectral windows with 480 channels each (0.168~\kms{} velocity resolution) and centered on \htwoco{} $\rm 3_{0,3} - 2_{0,2}$, and \htwoco{} $ 3_{2,2} - 2_{2,1}$, \htwoco{} $\rm 3_{2,1}-2_{2,0}$ and SO $\rm 6_5-5_4$. The third baseband was configured with a 0.94~GHz spectral window (1920 channels, 1.25~\kms) centered on $^{12}$CO 2--1. Finally, the fourth baseband contains a 1.875~GHz continuum band centered at 233.0~GHz with 1920 channels.

The data were reduced using the ALMA calibration pipeline within CASA 5.4.0 \citep{McMullin07}. In order to increase the signal-to-noise ratio of the continuum and spectral lines, we performed self-calibration on the continuum. We performed two rounds of phase-only self-calibration. The first round used solution intervals that encompassed the length of an entire on-source scan, then the second round utilized the 6.05 s solution interval, corresponding to a single integration. The phase solutions from the continuum self-calibration were also applied to the spectral line bands. The resultant rms noise in the 1.3~mm continuum was $\sim$0.13 \mjypbm. The continuum and spectral line data were imaged using the {\it tclean} task within {CASA 5.4.0} with Briggs weighting and a robust parameter of 0.5. The beam size of the continuum is 0\farcs24$\times$0\farcs21 (103~au $\times$ 90~au). And the observation can recover fluxes at spatial scales up to 2\farcs{9}. In \autoref{table:line_info} we list the information of lines that are used in this study, including both the lines mentioned above and those identified in the continuum spectral window (note that this is not an exhaustive list of spectral lines contained with in the dataset, also see \autoref{sec:app_lines}).

\section{Results}\label{sec:res}

\begin{figure*}[ht!]
\epsscale{1.1}\plotone{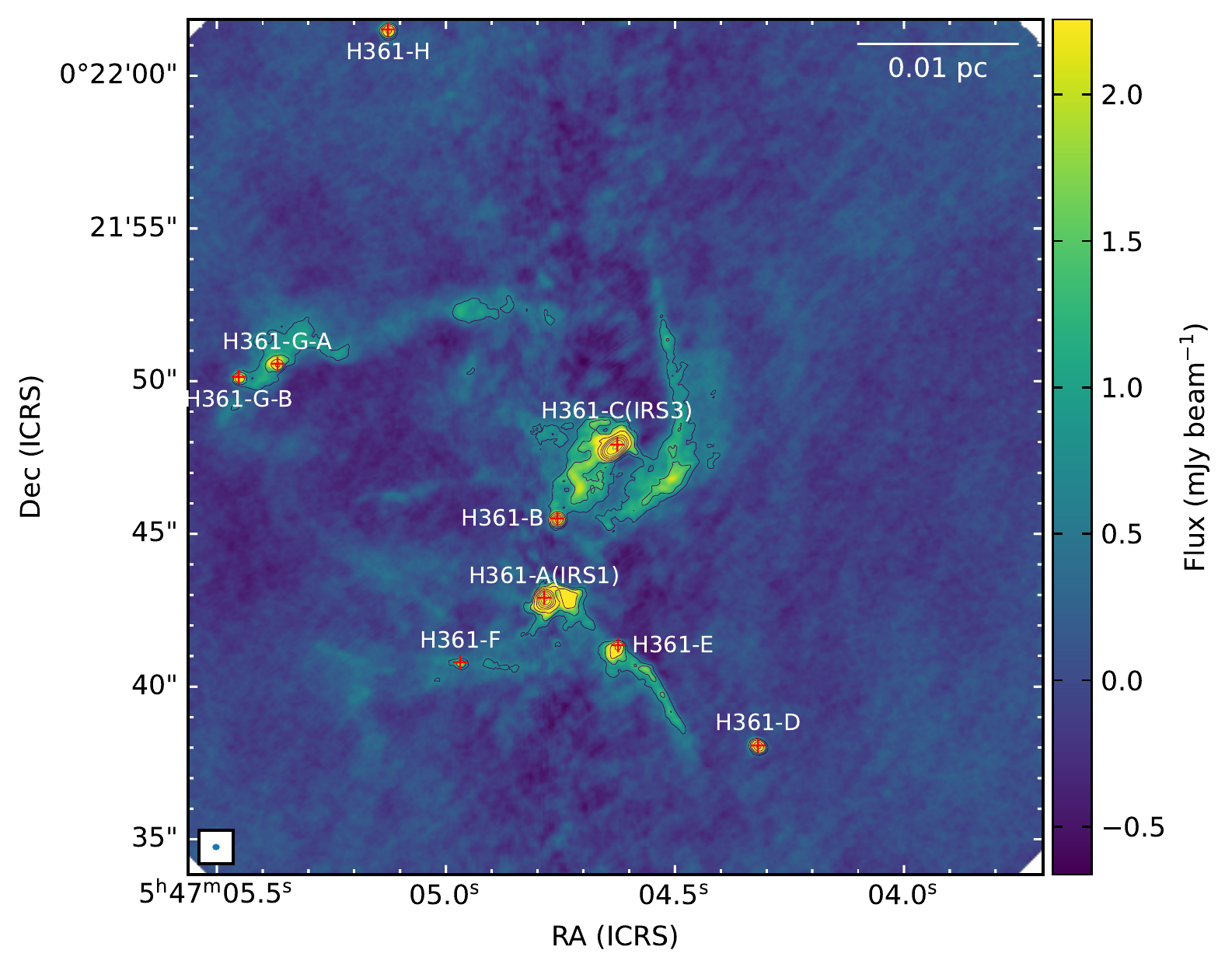}
\caption{Overview of the NGC~2071~IR region. The 1.3~mm continuum is shown in colorscale and contours. The contours levels are (5, 10, 20, 50, 100, 200, 400, 800) $\times \sigma$, where $\sigma$ = 0.13 \mjypbm. The position of identified protostar sources in \citet{Tobin20} are marked by red crosses and labeled in white text. The designation ``HOPS-361'' is abbreviated to ``H361''. The beam size is 0\farcs{24} $\times$ 0\farcs{21} as shown in the bottom left corner.\label{fig:overview}
}
\end{figure*}

\subsection{ALMA and VLA Continuum Images}
\label{sec:cont}

\autoref{fig:overview} illustrates the 1.3~mm continuum map of the NGC~2071~IR region. The overall SED of this region at shorter wavelengths has been studied in \citet{Furlan16} with photometry data from 2MASS, {\it Spitzer} and {\it Herschel} (i.e., the source HOPS-361 following their designation). \citet{Tobin20} identified 8 protostar systems based on high resolution ALMA 0.87~mm and VLA 9~mm observations. These sources, named from HOPS-361-A to HOPS-361-H, are labeled with red crosses in \autoref{fig:overview}. Five of them are associated with near-IR point sources \citep[i.e., IRS1, IRS2, IRS3, IRS4, IRS8,][]{Persson81,Walther93}. HOPS-361-G (IRS2) is known to be a binary system (HOPS-361-G-A, HOPS-361-G-B) separated by $\sim$1.4\arcsec ($\sim$580~au) \citep{Carrasco12}. HOPS-361-C (IRS3) and HOPS-361-E appear single in ALMA 0.87~mm continuum but are resolved to be close ($<$ 0\farcs{2} or 80 au) binary systems in the VLA 9~mm images. 

In the 1.3~mm map, these sources (HOPS-361-A to HOPS-361-H) are all detected at $>5\sigma$ level and exhibit compact dusty structures at 0\farcs{2} scale, which arise from their protostellar disks and inner envelopes. With self-calibration, our 1.3~mm map reaches a high dynamic range of $\sim$ 1000, and some weak extended structures have also been revealed. HOPS-361-C appears to be embedded in a larger dusty structure, which extends to the SE direction and connects with HOPS-361-B. There also appear to be a few filamentary features that are about 0.01--0.02~pc long: one originating from HOPS-361-E and extending to the SW direction, one extending from HOPS-361-G to the west, and another one spiraling around HOPS-361-C and extending to the north. The origin of these streamer features is not clear but is likely to be related with density enhancements shaped by complex gas motions on larger scales like ongoing infall \citep[e.g.,][]{Alves20}.

In \autoref{fig:YSO1} we present the ALMA (0.87~mm, 1.3~mm) and VLA (9~mm) continuum images towards these sources. We fit elliptical Gaussians to these protostellar sources using the {\it imfit} task in {CASA} to measure their positions, flux densities, and sizes in 1.3~mm, as listed in \autoref{table:YSOb} and \autoref{table:YSOa}. The fluxes and sizes in 0.87~mm and 9~mm from \citet{Tobin20}, which are measured with the same method, are also listed for comparison. In \autoref{sec:app_sed} we analyse the SED of these sources from 0.87~mm to 20~cm. Most sources show an SED consistent with free–free thermal emission at centimeter wavelengths and thermal dust emission at millimeter. 

We use the flux densities at 0.87 mm and 1.3 mm to calculate the mass of the material surrounding the protostars, assuming that the emission purely comes from optically thin isothermal dust emission, enabling us to use the equation

\begin{equation}
    M_{\rm dust} = \frac{D^2F_\nu}{\kappa_\nu B_\nu (T_{\rm dust})}.
\end{equation}

In this equation, $D$ is the distance, $F_\nu$ is the observed flux density, $B_\nu$ is the Planck function, $T_{\rm dust}$ is the dust temperature and $\kappa_\nu$ is the dust opacity at the observed wavelength. We adopt $\kappa_{\rm 0.87 mm}$ = 1.84~$\rm cm^2g^{-1}$ and $\kappa_{\rm 1.3 mm}$ = 0.899~$\rm cm^2g^{-1}$ from \citet{Ossenkopf94} (thin ice mantles, $\rm 10^6~cm^{-3}$ density). We multiply the calculated dust mass by 100, assuming a dust-to-gas mass ratio of 1:100 \citep{Bohlin78}, to obtain the gas mass. The average dust temperature we adopt for a protostellar system is given by

\begin{equation}
    T_{\rm dust} = T_0 \left( \frac{L}{L_\odot} \right) ^{0.25},
\end{equation}
where $T_{0}$ = 43 K. The average dust temperature of 43~K is reasonable for a $\sim$1 $L_\odot$ protostar at a radius of $\sim$50~au \citep{Whitney03,Tobin13}. For the luminosity of this region, \citet{Furlan16} has estimated a total $L_{\rm bol}$ of 478~$L_\odot$ in an aperture encompassing both IRS1 and IRS3. For simplicity we calculate the relative $L_{\rm bol}$ ratios among IRS1, IRS2 and IRS3 based on the SOFIA 37.1~$\mu$m image, which is the longest infrared wavelength for which we can still resolve the three components (see \autoref{sec:sed}). Thus the $L_{\rm bol}$ is 368~$L_\odot$, 25~$L_\odot$, and 85~$L_\odot$ for IRS1, IRS2, and IRS3, respectively. For other sources without a measured $L_{\rm bol}$ we adopt 1~$L_\odot$. The $L_{\rm bol}$, $T_{\rm bol}$, derived masses, as well as other available identifiers of the protostars are listed in \autoref{table:YSOb}. The continuum emission from the protostars is likely to be partially optically thick; thus, the masses are likely lower limits, especially at 0.87~mm \citep[e.g.,][]{Reynolds21}.

\startlongtable
\begin{deluxetable*}{cccccccc}
\tabletypesize{\scriptsize}
\renewcommand{\arraystretch}{1.0}
\tablecaption{Source properties of protostars in NGC~2071~IR\label{table:YSOb}}
\tablehead{
\colhead{Source} & \colhead{Other identifiers} & \colhead{R.A.$^a$} & \colhead{Decl.$^a$} & \colhead{$L_{\rm bol}$} & \colhead{$T_{\rm dust}$} &\colhead{$M_{\rm 1.3mm}$}  & \colhead{$M_{\rm 0.87mm}$}\\
\colhead{} & \colhead{} & \colhead{(J2000)} & \colhead{(J2000)} & \colhead{($L_{\odot}$)} & \colhead{(K)} & \colhead{($M_\odot$)} & \colhead{($M_\odot$)} 
}
\startdata
HOPS-361-A & IRS1 & 5:47:4.784 & 0:21:42.85     & 368 & 188 &  0.0690 $\pm$ 0.0017 & 0.0611 $\pm$ 0.0006\\
HOPS-361-B & VLA1 & 5:47:4.755 &0:21:45.45     & 1 & 43 &  0.0645 $\pm$ 0.0012 & 0.0486 $\pm$ 0.0013\\
HOPS-361-C & IRS3 & 5:47:4.631&0:21:47.82     & 85 & 131 &  0.1750 $\pm$ 0.0039 & 0.1309 $\pm$ 0.0014\\
HOPS-361-D & IRS8 & 5:47:4.317&0:21:38.03     & 1 & 43 &  0.0360 $\pm$ 0.0004 & 0.0367 $\pm$ 0.0022\\
HOPS-361-E & - & 5:47:4.623&0:21:41.30     & 1 & 43 &  0.0392 $\pm$ 0.0053 & 0.0151 $\pm$ 0.0022\\
HOPS-361-F & - & 5:47:4.967&0:21:40.74     & 1 & 43 &  0.0080 $\pm$ 0.0009 & 0.0034 $\pm$ 0.0006\\
HOPS-361-G-A & IRS2A & 5:47:5.367&0:21:50.51     & 25 & 96 &  0.0115 $\pm$ 0.0015 & 0.0065 $\pm$ 0.0003\\
HOPS-361-G-B & IRS2B & 5:47:5.451&0:21:50.08     & 25 & 96 &  0.0061 $\pm$ 0.0006 & 0.0039 $\pm$ 0.0003\\
HOPS-361-H & IRS4 & 5:47:5.125&0:22:1.46     & 1 & 43 &  0.0223 $\pm$ 0.0006 & -\\
\enddata
\tablenotetext{a}{Positions measured from the 1.3~mm continuum by 2-D Gaussian fits. }
\end{deluxetable*}

\startlongtable
\begin{deluxetable*}{cccccccccc}
\tabletypesize{\scriptsize}
\renewcommand{\arraystretch}{1.0}
\tablecaption{Flux intensities and sizes of protostars in NGC~2071~IR$^{a}$ \label{table:YSOa}}
\tablehead{
\colhead{Source} & \colhead{$F_\nu$(1.3mm)} & \colhead{Size(1.3mm)} & \colhead{PA(1.3mm)} & \colhead{$F_\nu$(0.87mm)} & \colhead{Size(0.87mm)} & \colhead{PA(0.87mm)} & \colhead{$F_\nu$(9mm)} & \colhead{Size(9mm)} & \colhead{PA(9mm)} \\
& \colhead{(mJy)} & (arcsec) & {(degree)} & \colhead{(mJy)} & (arcsec) & {(degree} & \colhead{(mJy)} & (arcsec) & {(degree)} 
}
\startdata
HOPS-361-A &208.0   &0.24$\times$0.14 &12.0 &606.4 & 0.26$\times$ 0.15  & 14.4 & 5.6 &0.10$\times$0.07 & 64.4 \\
HOPS-361-B &40.3    &0.08$\times$0.06 &174.4 &94.4   & 0.06$\times$ 0.05 & 174.4& 1.9 &0.07$\times$0.04 &55.5  \\
HOPS-361-C-A$^{b}$ &362.5 &0.48$\times$0.19 &129.5 &883.4 & 0.47$\times$ 0.19  & 130.1 & 2.3 &0.14$\times$0.04 & 6.7 \\
HOPS-361-C-B$^{b}$ &  -    &         -        &- &- & - & -                      & 0.2 & point & point \\
HOPS-361-D &22.5    &0.19$\times$0.10 &44.2 &71.3& 0.17$\times$ 0.09& 43.1& 0.4 &0.09$\times$0.06 &50.9 \\
HOPS-361-E &15.3    &0.30$\times$0.19 &147.0 &29.4& 0.15$\times$ 0.11 &111.9  & 0.2&point &point  \\
HOPS-361-F &4.2     &0.26$\times$0.11 &79.0 &6.6&point &point &0.1 & point & point \\
HOPS-361-G-A &22.6  &0.53$\times$0.35 &136.0 &31.6 &point &point & 0.4 &0.09 $\times$0.06 & 44.3 \\
HOPS-361-G-B &9.3   &0.22$\times$0.19 &134.0 &19.0 &point &point & 0.3 &0.07$\times$0.04 & 13.3 \\
HOPS-361-H &13.9    &0.20$\times$0.19 &152.0 &-$^{c}$ &-$^{c}$ &-$^{c}$ &0.6 &  0.08 $\times$0.06 & 18.0  \\
\enddata
\tablenotetext{a}{The sizes and P.A. are the FWHM and P.A. of deconvolved Gaussian components. }
\tablenotetext{b}{For HOPS-361-C (IRS3) the two components are resolved only in 9~mm. The listed fluxes and sizes in 0.87~mm and 1.3~mm refer to the properties of the circumbinary disk. }
\tablenotetext{c}{HOPS-361-H is not covered in the 0.87~mm observation.}
\end{deluxetable*}

\begin{figure*}[ht!]
\epsscale{1.1}\plotone{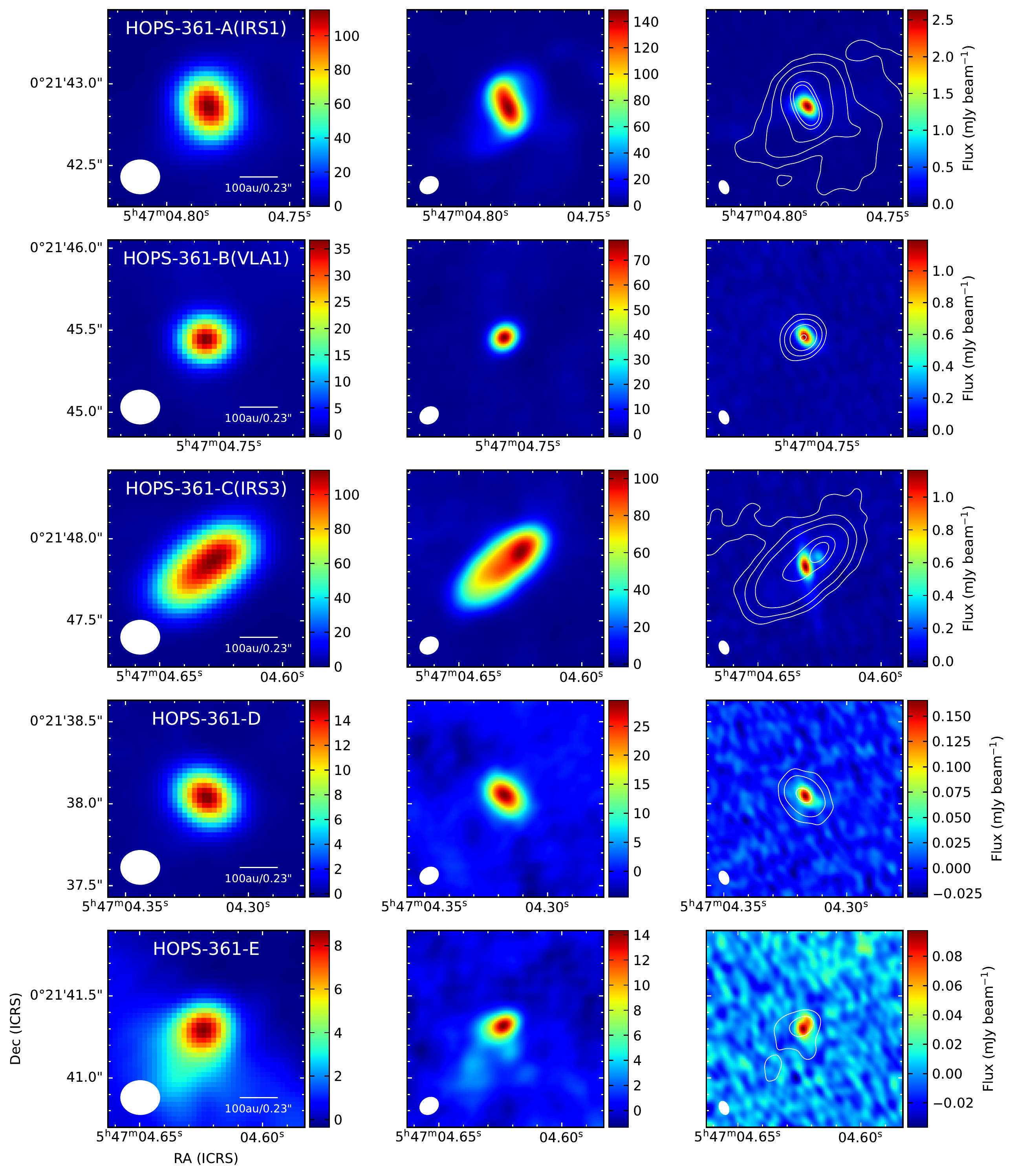}
\caption{Continuum images at 1.3 mm (left), 0.87 mm (center) and 9 mm (right) of the protostars in the NGC~2071~IR region. For the 9~mm images we overplot the 0.87~mm continuum in white contours for comparison. The contours levels are (5, 15, 45, 135, 170) $\times $ 0.55~\mjypbm. The beam sizes are 0\farcs{24} $\times$ 0\farcs{21} (104~au $\times$ 91~au)  for 1.3~mm, 0\farcs{13} $\times$ 0\farcs{10} (56~au $\times$43~au) for 0.87~mm, and 0\farcs{09} $\times$0\farcs{06} (39~au $\times$ 26~au) for 9~mm, as illustrated in the bottom left corner of each panel.
}
\label{fig:YSO1}
\end{figure*}

\addtocounter{figure}{-1}
\begin{figure*}[ht!]
\epsscale{1.1}\plotone{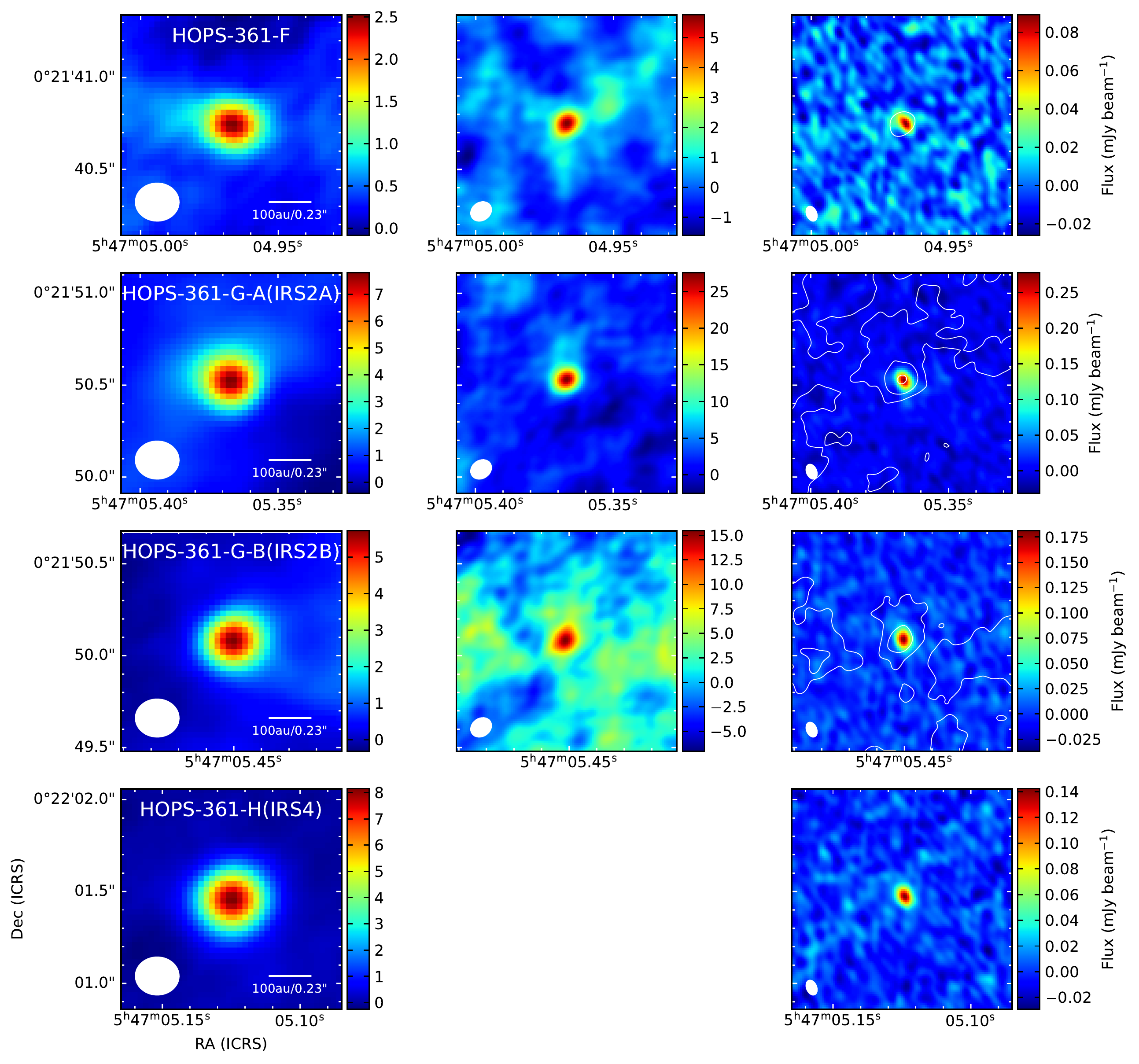}
\caption{Continued. HOPS-361-H is not covered in the FOV of the 0.87~mm observation.\label{fig:YSO2}
}
\end{figure*}

\begin{figure*}[ht!]
\epsscale{1.2}\plotone{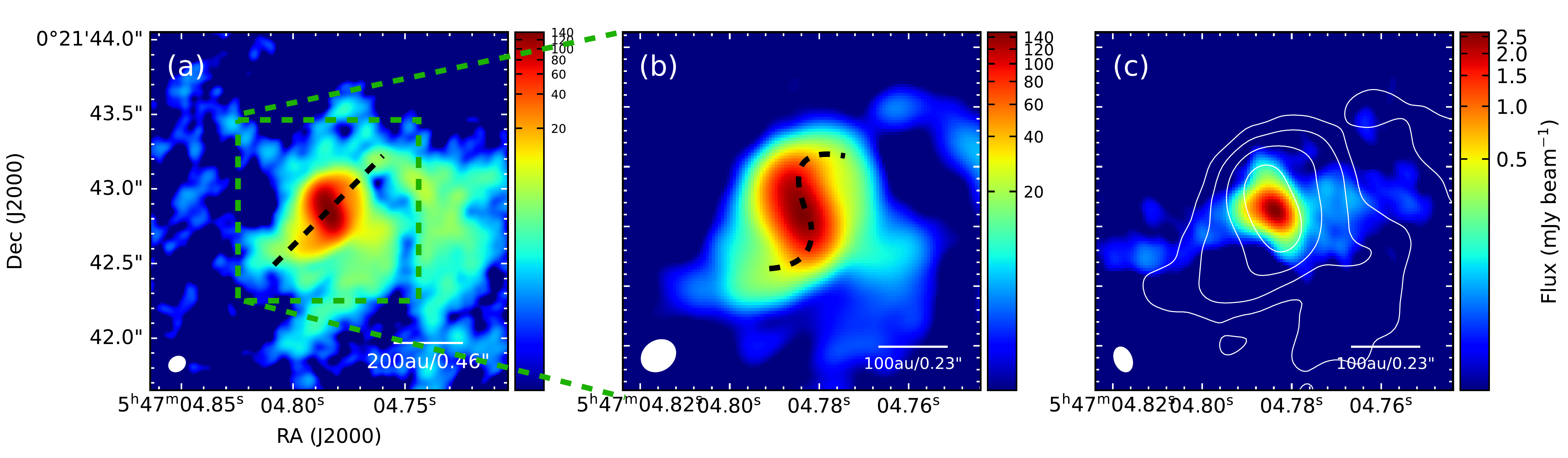}
\caption{0.87~mm and 9~mm images of IRS1 shown in colorscale with a log stretch. (a) 0.87 mm continuum image of IRS1. The black dashed line indicate the disk orientation inferred from molecular line kinematics. (b) A zoom-in view of (a). The locations of possible spiral features are marked in dashed lines. (c) 9 mm continuum image of IRS1. The contours are 0.87 mm continuum with levels of (5, 15, 45, 135) $\times \sigma$ and $\sigma$ = 0.55~\mjypbm. The beam sizes are shown in the lower left corner.\label{fig:IRS1}}
\end{figure*}

\subsubsection{IRS1 and IRS3}

Among these sources, HOPS-361-A (hereafter IRS1) and HOPS-361-C (hereafter IRS3) have the strongest 1.3~mm continuum, with peak intensities of 152 and 114~\mjypbm, respectively. As stated in \autoref{sec:intro}, IRS1 and IRS3 are the dominant mid/far-IR luminosity contributors and also presumed driving sources of the large scale outflow \citep[e.g.,][]{Torrelles98,Eisloffel00}. 
As shown in \autoref{fig:YSO1} (see also \autoref{fig:IRS1}), IRS1 is partly resolved at 1.3~mm and no clear elongation is apparent. 
At 0.87~mm IRS1 appears better resolved. The inner brighter part of IRS1 (i.e., flux intensity above 60~\mjypbm) has a bar-like shape, with an elongation at P.A. of about 25\arcdeg. This elongated structure is further embedded in low level extended emission (below 60~\mjypbm{} but still above 20$\sigma$ = 11~\mjypbm). This weaker component is approximately elliptical and its major axis extends about 0\farcs{3} along the NW-SE direction, i.e., $\sim$ 110\arcdeg{} offset in orientation from the inner bright component. There are some hints of spiral-like bending features at the interface between the two components, and may further connect with larger scale spiral-like features extending to $\sim$1\arcsec\ (see \autoref{fig:IRS1}). These features have added to the complexity on inferring the configuration of the protostellar disk of IRS1. The kinematic information, which will be discussed in following sections, is more supportive of a disk major axis oriented in the NW-SE direction, i.e., consistent with the extended component. In this scenario, the inner bright component revealed in 0.87~mm appears as an unusual substructure of the protostellar disk. The deconvolved FWHM from Gaussian fit to the 0.87~mm continuum emission, which is dominated by the inner component, is 0\farcs{25}$\times$0\farcs{15} (108~au$\times$65~au).

The VLA 9~mm continuum, which traces both free-free emission and thermal dust emission, shows a distinctly different morphology with respect to the ALMA images. The 9~mm continuum emission of IRS1 appears as a marginally resolved condensation, which coincides well with the position of 0.87~mm flux peak. The 9~mm emission has a T-shape elongation, i.e., with the brighter part extending along the NE-SW direction (P.A. $\sim$ 25\arcdeg), and a fainter part extending slightly to the east (\autoref{fig:IRS1}). The NE-SW extension has a direction similar to the brighter part seen in 0.87~mm. In addition, some low level ($\sim$5$\sigma$) diffuse emission is also seen to the east and west of the central source, which extends as far as 0\farcs{5} (see \autoref{fig:IRS1}(c)). \citet{Trinidad09} reported radio knots ejected from IRS1 along the E-W direction (IRS1E, IRS1W) based on VLA 1.3~cm continuum. Comparing with their detections, the weak diffuse emission in the 9~mm could also trace a radio jet in the E-W direction. And it is likely that some of the previously detected radio knots, like IRS1W, have dissipated most of their energy and can no longer be detected, thus absent in our map, though the different surface brightness sensitivities in the two observations may hinder a conclusive interpretation.

On the other hand, IRS3 shows a clear disk at 0.87~mm and 1.3~mm. Gaussian fits to the continuum in both bands give a similar deconvolved size of 0.48\arcsec$\times$0.19\arcsec{} with a position angle of 130\arcdeg. Assuming that the Gaussian semi-major axis corresponds to the disk radius, it has a radius of 103~au, and the inclination can be estimated to be 67\arcdeg{} by assuming that it is a geometrically thin disk and then calculating the inverse cosine of the minor axis divided by the major axis. The flux distribution in 0.87~mm continuum appears asymmetric, with the emission peak offset by 0.16\arcsec($\sim$69 au) to the northwest compared with the geometric center. 
A similar asymmetry could be present at 1.3 mm, but this is less clear due to the lower spatial resolution.
Interestingly, in the center of the disk, the 9~mm continuum further reveals a binary system separated by 0\farcs{1} ($\sim$43~au). The two components (IRS3A, IRS3B) have a flux ratio of $\sim$10 at 9~mm, and the more luminous component, IRS3A, is coincident with the geometric center of the disk structure seen at 0.87~mm and 1.3~mm. IRS3B is located to the northwest of IRS3A and closer to the emission peak at 0.87~mm. IRS3B could be (at least partly) contributing to the asymmetric flux distribution seen at 0.87~mm via enhanced heating towards the surrounding dust/gas. While the detection of IRS3B is a point source, IRS3A is resolved and extends along the NE-SW direction to about 0\farcs{24} ($\sim$100 au) on both sides, with a position angle of $\sim$15\arcdeg. Extension from IRS3 in this direction has been reported in earlier VLA 1.3~cm and 3.6~cm observations, albeit with a lower resolution \citep{Carrasco12,Trinidad09}, and interpreted as a radio jet. 

\begin{figure*}[ht!]
\epsscale{1.15}\plotone{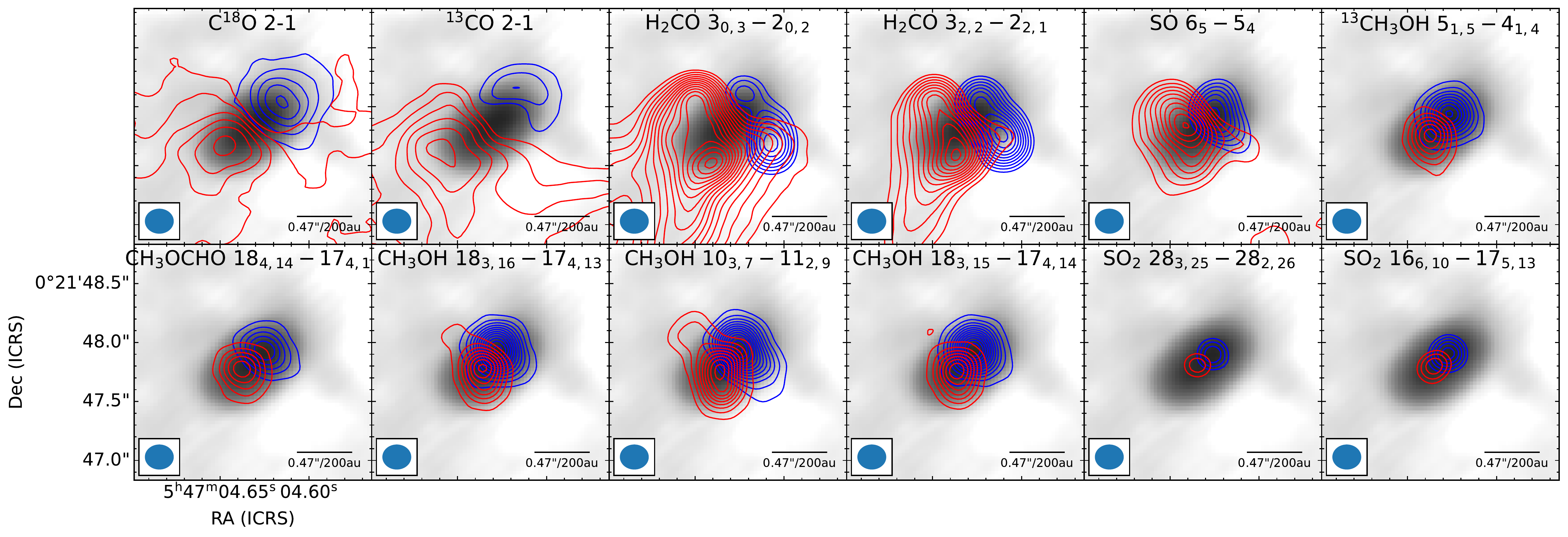}
\caption{Integrated intensity maps of spectral lines toward IRS3 overlaid on the 1.3~mm continuum (gray scale). The transitions are marked on top of the panel. The integrated intensity maps are separated into blueshifted velocities at 3 -- 8~\kms{} and redshifted velocities at 11 -- 16~\kms{}, and plotted with blue and red contours, respectively. For transitions from \thirteenco, \htwoco{} and SO the contours start from 40$\sigma$ in steps of 10$\sigma$. For other lines the contours start at 10$\sigma$ and increase on 10$\sigma$ intervals. $\sigma$ = $\sqrt{N_{\rm chan}}\times \sigma_{\rm chan} \times \Delta v$, where $\sigma_{\rm chan}$ is the rms noise per channel and $\Delta v$ is the velocity resolution, as listed in \autoref{table:line_info}.\label{fig:lines_IRS3}
}
\end{figure*}

\begin{figure*}[ht!]
\epsscale{1.15}\plotone{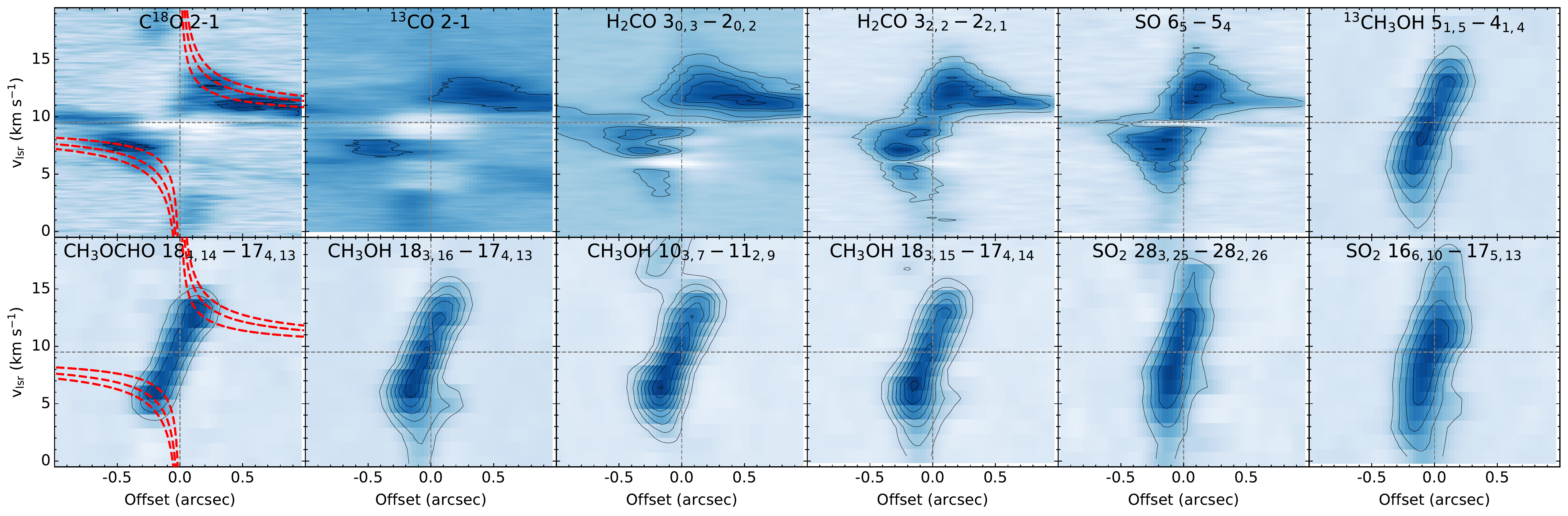}
\caption{PV diagrams of spectral lines toward IRS3, cut along the disk midplane, i.e., along P.A. $\sim$130\arcdeg, as determined from the 2D gaussian fit of the dust continuum. The transitions are marked on top of each panel. The contours start from 10$\sigma$ and increase in steps of 10$\sigma$. The vertical line marks the reference point, i.e., the source position determined from the 2D Gaussian fit of the 1.3~mm continuum; the horizontal line indicates the systemic velocity. In the left two panels, we overplot in dashed red lines the Keplerian rotation curves for central masses of 1, 2 and 3~\msun{} as a reference. An inclination of 67\arcdeg, as estimated from the dust continuum, is assumed. } \label{fig:pv_IRS3}
\end{figure*}

\begin{figure*}[ht!]
\epsscale{1.15}\plotone{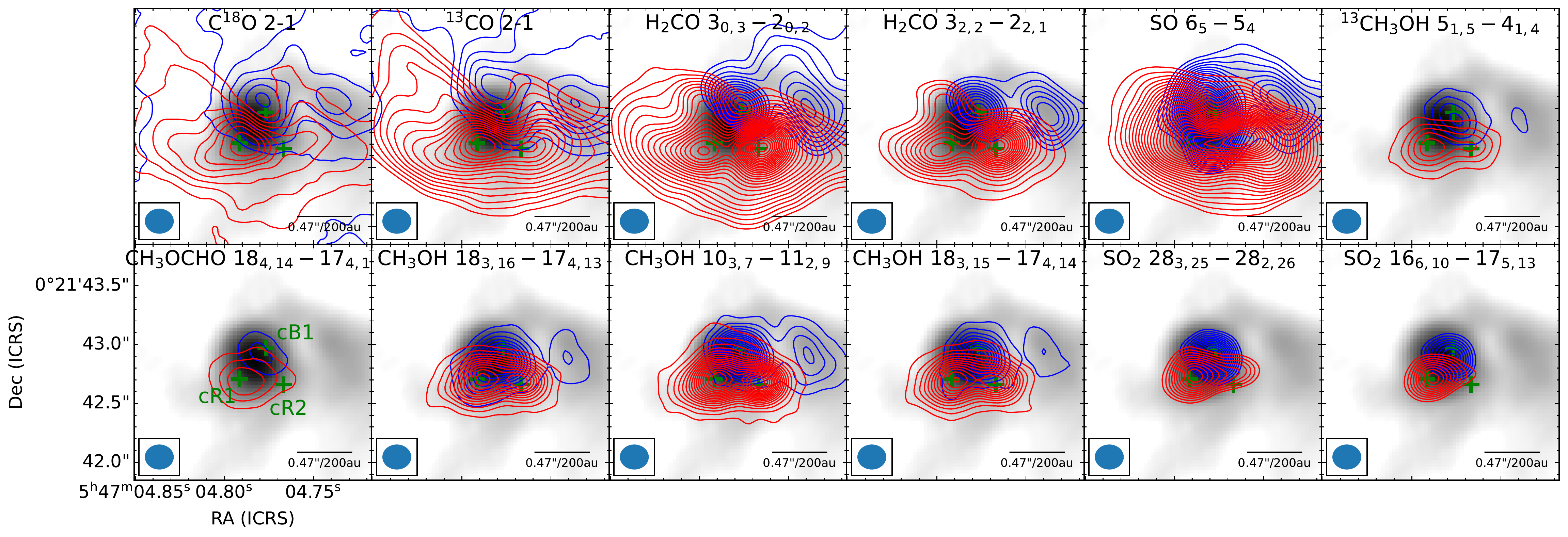}
\caption{Same as \autoref{fig:lines_IRS3} but for IRS1. The integrated intensity maps are separated into blueshifted velocities at 3 -- 8~\kms{} and redshifted velocities at 10 -- 15~\kms. For transitions from \thirteenco, \htwoco{} and SO the contours start from 40$\sigma$ in steps of 10$\sigma$. For other lines the contours start at 10$\sigma$ and increase on 10$\sigma$ intervals. We have labeled the positions of cB1, cR1 and cR2 in green crosses (see text for more details).}\label{fig:lines_IRS1}
\end{figure*}

\begin{figure*}[ht!]
\epsscale{1.15}\plotone{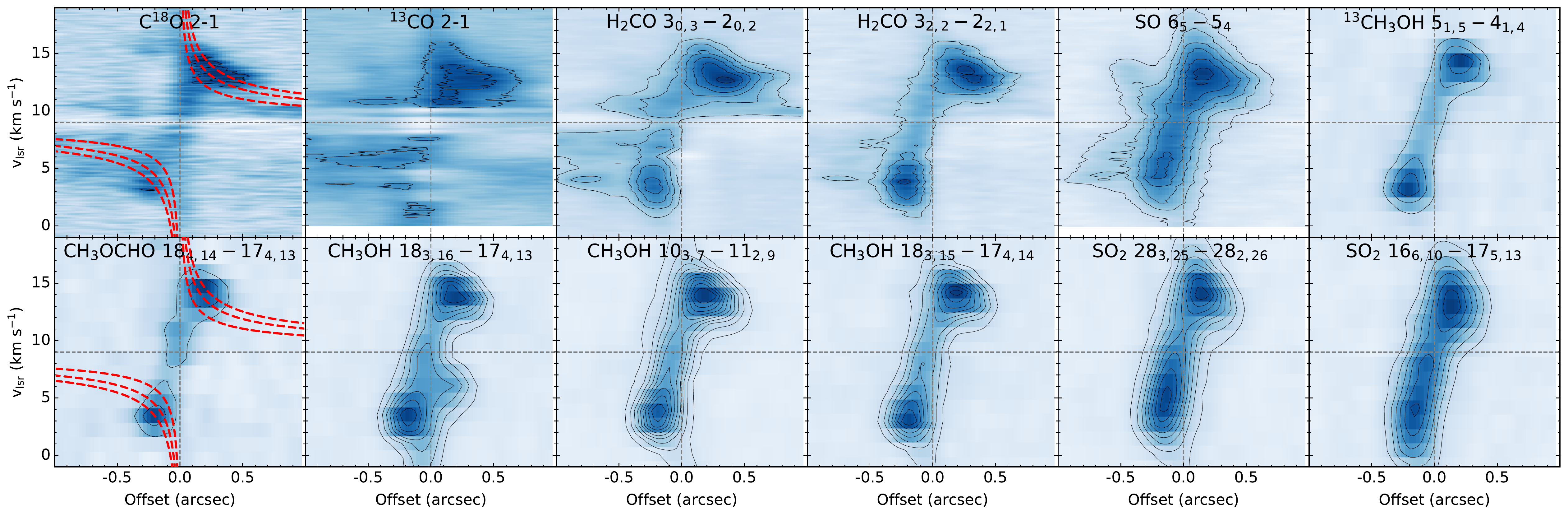}
\caption{Same as \autoref{fig:pv_IRS3} but for IRS1. The PV diagram is cut along P.A. $\sim$135\arcdeg{} (see \autoref{fig:IRS1} and text for details). In the left two panels, we overplot in dashed red lines the Keplerian rotation curves for central masses of 2, 4 and 6~\msun{} as a reference. An inclination of 45\arcdeg{} is assumed.} \label{fig:pv_IRS1}
\end{figure*}

\subsection{Molecular Line detections}\label{sec:lines}

We have detected a series of molecular lines associated with the protostellar disks for both IRS1 and IRS3, including transitions from \ceighteeno, \thirteenco, \htwoco, \methanol, \thirteenmethanol{}, and  \sotwo. In addition to these lines, we have also detected abundant lines in the continuum spectral window in band 6. Detailed modelling with \textsc{xclass} \citep{Moller17} suggests most of these lines arise from organic molecules like $\rm CH_3OCHO$ and $\rm NH_2CHO$, etc (see \autoref{sec:app_lines}). For all the lines the spectra averaged over the disk usually exhibit a double peak profile that is most likely arising from disk rotation (e.g, see the the COM lines presented in \autoref{sec:app_lines}). We did not find infalling signatures like red-shifted absorption in lines that are likely optically thick, i.e., from \ceighteeno, \thirteenco, and \htwoco.

\subsubsection{IRS3}
\autoref{fig:lines_IRS3} presents the integrated intensity map of a selected number of spectral lines toward IRS3. Line emission integrated over two velocity intervals (1.5~\kms$< |v - v_{\rm sys}| <$ 6.5~\kms, relative to a systemic velocity $v_{\rm sys} \approx$ 9.5~\kms) is shown in blue and red, respectively. Almost all lines exhibit a clear velocity gradient along the major axis of the millimeter continuum, with emission transitioning from blueshifted in the northwest to redshifted in the southeast. This monotonic velocity transition and its correspondence with the dust continuum are strongly indicative of a Keplerian rotating disk. \autoref{fig:lines_IRS3} also reveals the difference in spatial distribution of line emission from different molecules. Species including \ceighteeno, \thirteenco, \htwoco, and \so{} exhibit strong emission beyond the disk boundary defined by 1.3~mm/0.87~mm dust continuum. The line emission from \thirteenmethanol, \methanol, \sotwo{}, and other organic molecules are more spatially compact, i.e., within 0\farcs{25} ($\sim$ 108~au) from the center. Hereafter we refer to the two groups of lines with distinct morphology as group A and group B, i.e., group A lines include those from \ceighteeno, \thirteenco, \htwoco{} and \so, while groups B lines include transitions from \methanol, \sotwo, \thirteenmethanol, as well as other complex organic molecules ($\rm CH_3OCHO$ is shown here as an example).

These two categories are better illustrated in the PV diagram extracted along the major axis of the dust continuum, as shown in \autoref{fig:pv_IRS3}. 
The group A lines, i.e., \ceighteeno, \htwoco, \so, etc, show bright emission peaks in the first and third quadrant. For \thirteenco{} and \htwoco{} the detection in first quadrant (i.e., blueshifted emission) is stronger. For all species in group A the detection close to the systemic velocity is relatively weak, possible due to self-absorption of cold gas along line of sight and/or spatial filtering of the extended emission from the ambient cloud in interferometric observations, especially for \ceighteeno{} and \thirteenco. 
In contrast with the group A, lines in the group B appear as a continuous linear feature crossing the first and third quadrant, and no low velocity emission extending beyond 0\farcs{5} ($\sim$215~au) is apparent. The linear feature is consistent with a velocity gradient around 20~\kms~arcsec${^{-1}}$, or 0.046 ~\kms~au${^{-1}}$, and the intensity distribution across it is relatively uniform. The spatial extents of these lines line up well with the disk boundary inferred from 1.3~mm/0.87~mm continuum. Nevertheless, the outer emission edges on the PV diagram have a convex shape, 
in contrast with the expectation for a Keplerian disk, but we will show in \autoref{sec:kine} that this is mainly due to the limited spatial resolution. 
Our band 6 observation has a resolution of $\sim$0\farcs{25} (108 au), comparable with half of the major axis of dust continuum, so the detailed PV structures have been smoothed out in this plot, especially for the group B lines. 
The different spatial and kinematic distributions in the two groups are likely to be reflecting different excitation conditions required for different transitions, e.g., group A and B lines have systematically different upper energy levels (see \autoref{table:line_info}).

\subsubsection{IRS1}
The kinematics of IRS1 are more difficult to infer since we do not have clear knowledge about the disk orientation from dust continuum. \autoref{fig:lines_IRS1} presents the integrated intensity map of spectral lines toward IRS1. There is an extended blueshifted structure, about 0\farcs{6} to the west of IRS1, seen in \ceighteeno, \thirteenco, \htwoco, \so{} and \methanol. This feature is not associated with the inner disk of IRS1 and should be tracing the extended emission adjacent to it (see \autoref{fig:overview} or \autoref{fig:IRS1}). Interestingly, for most species there appear to be one blueshifted clump (cB1) and two redshifted clumps (cR1, cR2) associated with IRS1. This is most clear for \htwoco, \so{} and \methanol. For an individual rotating disk, one would expect a monotonic velocity gradient along the major axis, as observed for IRS3. Here we think cB1 and cR1 are tracing the protostellar disk, while the third gas clump cR2 is a separate structure that is not associated with IRS1 disk, for the following reasons. Firstly, the position of cR2 is more spatially offset from the emission peak of dust continuum compared with cB1 and cR1. If cB1 and cR2 are tracing the gas rotation on both sides of a disk, then the inferred position of a disk will disagree with that traced by dust continuum. Secondly, not all the lines exhibit clear detection at the position of cR2. cR2 clump is absent in organic molecules like $\rm CH_3OCHO$ and for \sotwo{} and \thirteenmethanol{} only some weak extension from cR1 towards cR2 is seen. Again, this is in contrast with the expectation that cR2 is tracing one side of the disk since similar chemical/excitation properties are generally expected for both sides of a disk. 
Therefore the IRS1 disk, as traced by cB1 and cR1, is oriented in the NW-SE direction. The other peak cR2 is likely a shock knot excited by the ejection from the protostar.

\autoref{fig:pv_IRS1} illustrates the PV diagram extracted along the the inferred disk orientation of IRS1, i.e., following the direction of P.A. $\sim$ 135\arcdeg. Similar to IRS3, there are two categories of molecular lines: \ceighteeno, \thirteenco, \htwoco{} and \so{} show low velocity emission extending beyond 0\farcs{5}, while \thirteenmethanol, \methanol, \sotwo{} and other organic molecules are exclusively tracing the inner disk. But different from the prototypical case of IRS3, for the lines in the group B, the linear feature is composed of two separate emission peaks in the first and third quadrant, instead of a more continuous distribution. In addition, the PV diagram of IRS1 appears more asymmetric against the origin, in terms of both the intensity and the shape. This deviation from symmetric kinematics may arise from an imperfect determination of the disk position/orientation, or confusion by other mechanisms, like ejection, or hidden multiplicity inside IRS1.

\begin{figure*}[ht!]
\epsscale{1.0}\plotone{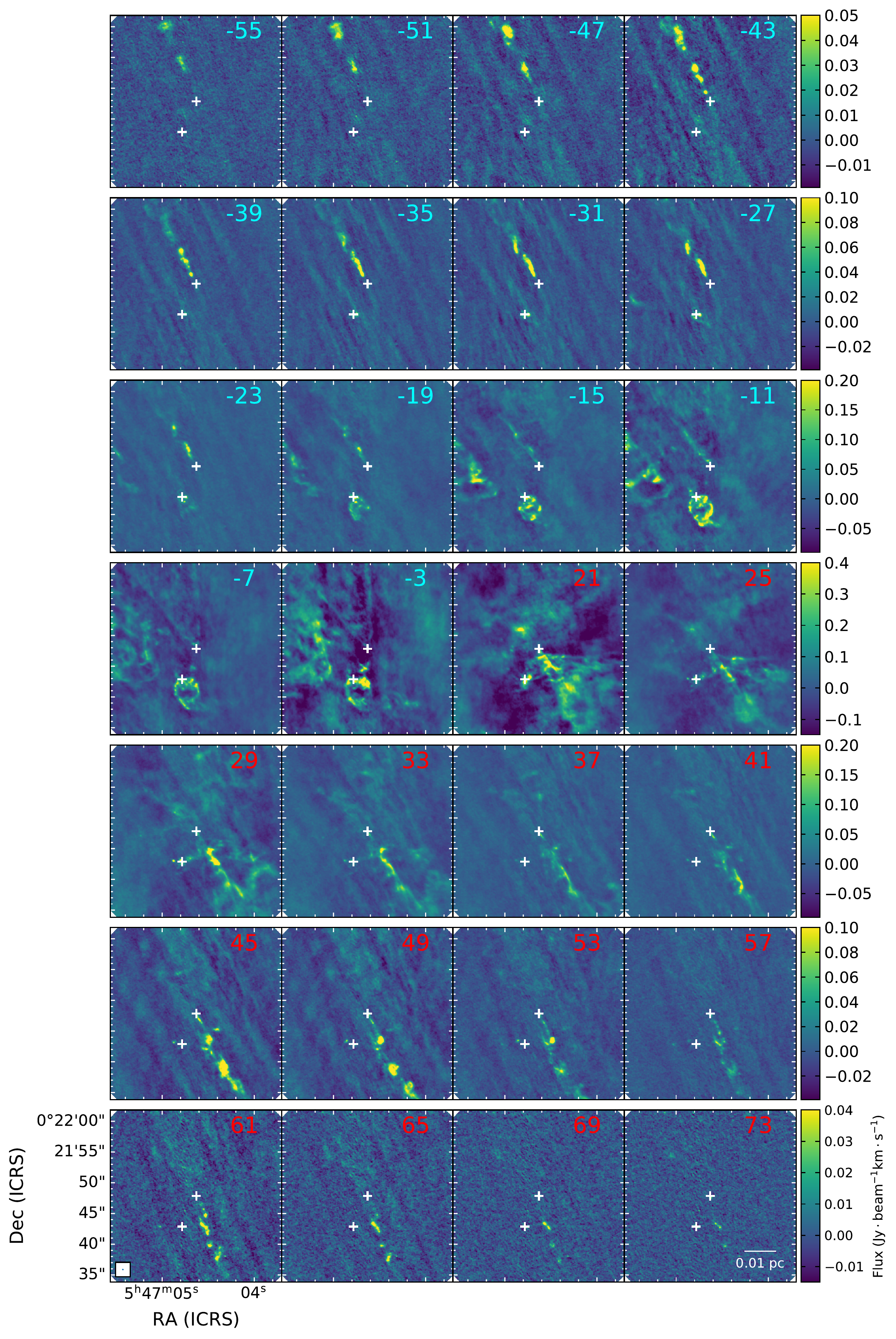}
\caption{CO intensity map integrated over every 4 \kms{} from $-$55 to $-$3~\kms, and 21~\kms{} to 73~\kms{} shown in colorscale. The center velocity of each panel is marked on the top right corner, in unit of \kms, with the blueshifted and redshifted velocities shown in blue and red text, respectively. The positions of IRS1 and IRS3 are indicated by white crosses. A scalebar is given in the bottom right panel.\label{fig:outflow_chan}
}

\end{figure*}

\begin{figure*}[ht!]
\epsscale{0.93}\plotone{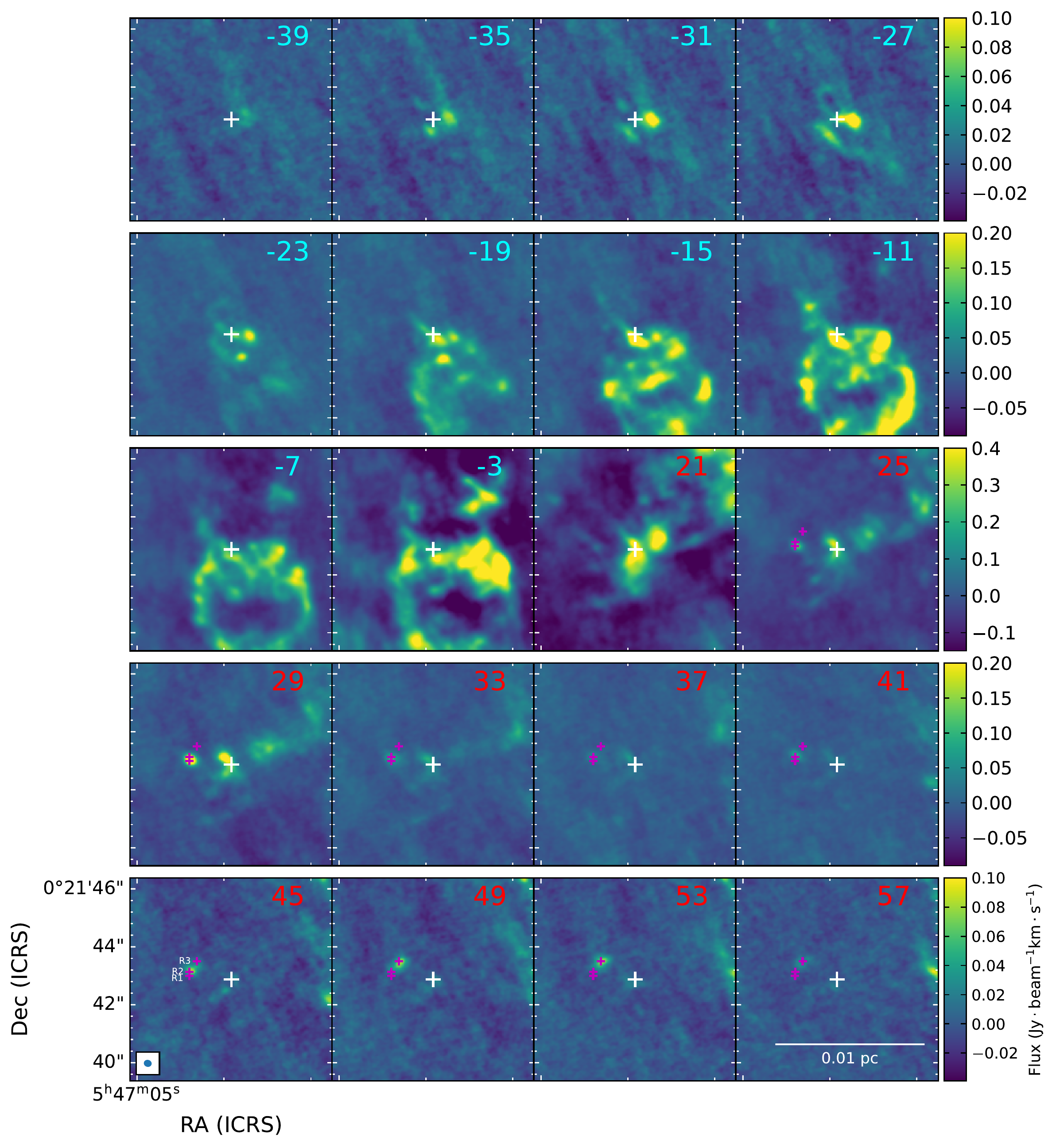}
\caption{Same as \autoref{fig:outflow_chan} but a zoom-in view for IRS1. We focus on CO intensity maps at velocities from $-39$ to 57~\kms, for which the outflows associated with IRS1 are more prominent. The positions of three redshifted clumps are marked in magenta crosses and labeled in the lower left panel (see text for more details). \label{fig:outflow_chan_IRS1}
}
\end{figure*}

\begin{figure*}[ht!]
\epsscale{0.7}\plotone{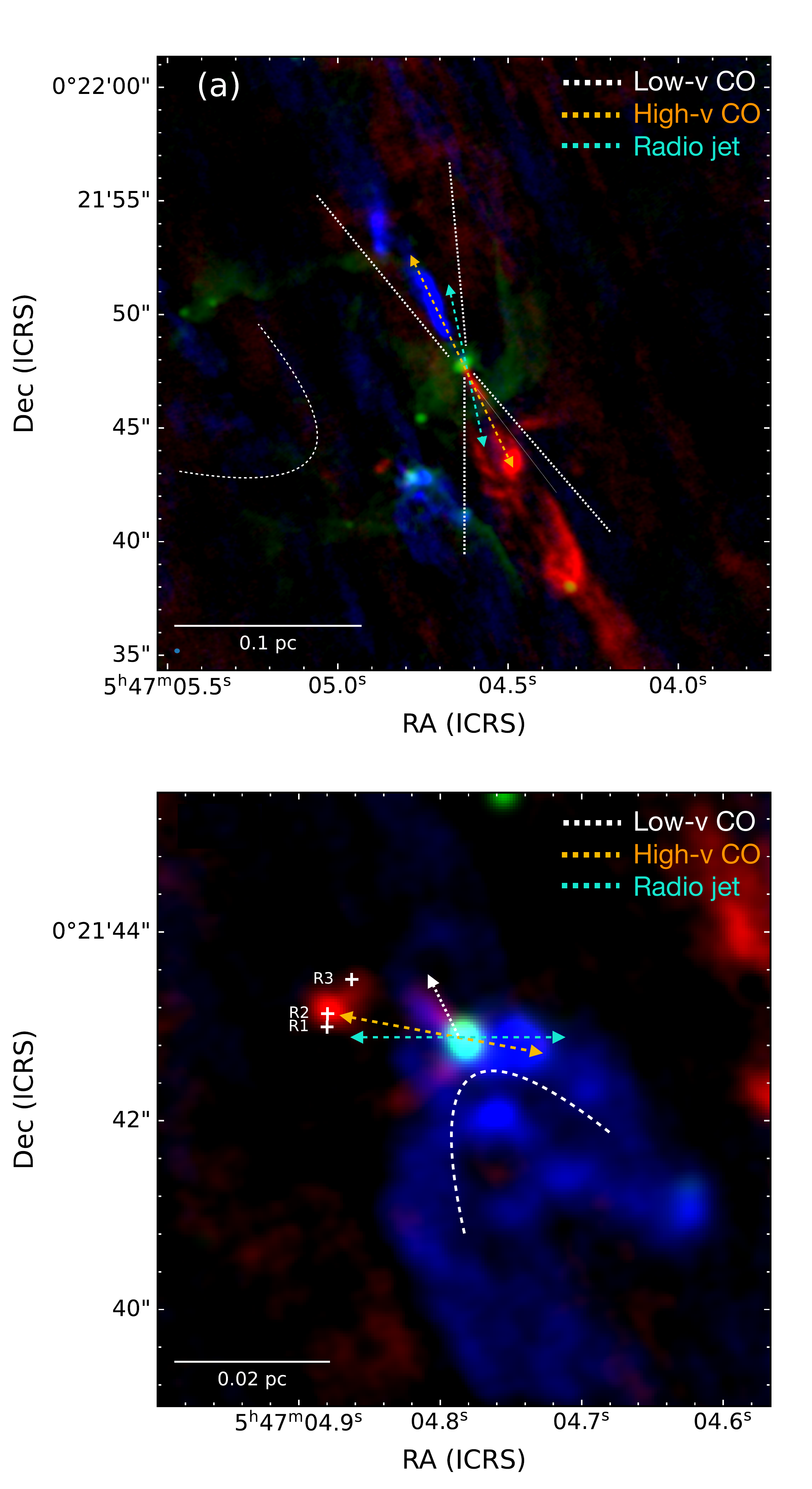}
\caption{ {\it (a)} An overview plot of the CO outflow detections associated with IRS3. The plot is overlaid on a three color image made with integrated blueshifted and redshifted CO emission, as well as the 1.3~mm continuum (in green).  The blue and red colorscales represent CO emission integrated over ($-$51, $-16$)~\kms{} and (34, 69)~\kms, respectively. The high velocity CO jet is indicated with a dashed brown line, while the CO cavity seen in low velocities are indicated with a dashed white line. The CO cavity to the east is likely associated with IRS1. Here the high velocity and low velocity components are roughly divided using a threshold of $|v-v_{\rm sys}|$ = 25~\kms.
The direction of radio jet, marked in a cyan dashed line, is determined from our 9~mm map \citep[see also][]{Trinidad09,Carrasco12}. {\it (b)} Same as {\it (a)} but a zoom-in view for IRS1. The blue and red colorscales represent CO emission integrated over ($-$31, $-16$)~\kms and (34, 49)~\kms, respectively. For IRS1, the dashed white lines indicate a bubble-like wide angle blueshifited outflow lobe identified in this work, while the dashed brown lines indicate the blueshifted and redshifted CO clumps seen at higher velocities (see text for more details). And the cyan dashed line indicates the approximate direction of the radio ejection reported in \citet{Trinidad09}. The positions of three redshifted clumps are marked in white crosses.\label{fig:outflow_sum}}
\end{figure*}

\subsection{Outflows in $^{12}CO$}
\label{sec:outflow}
Our observations also allow for a search for protostellar outflows associated with IRS1 and IRS3 via the $^{12}$CO 2--1 data. \autoref{fig:outflow_chan} presents the channel map of CO 2--1 integrated every 4~\kms{} from $-$55 to 73~\kms. The systemic velocities of IRS1 and IRS3 are around 9.0 and 9.5~\kms, respectively (see \autoref{sec:kine}). A jet-like outflow can be clearly seen in channels from $-$55 to $-$19~\kms{} for the blueshifted lobe, and from 29 to 73~\kms{} for the redshifted lobe. This jet is symmetrically distributed against IRS3 and extends to at least $\sim$ 0.02~pc long on both sides. The jet is approximately perpendicular to the IRS3 disk in the map.
The jet has an extremely high velocity, i.e., a maximum LOS velocity of $\sim$ 70~\kms{} relative to IRS3, or a true velocity of $\sim$ 150~\kms{} after correcting for an inclination angle of 67\arcdeg, as estimated in \autoref{sec:cont}. Overall the jet has a clumpy appearance in most panels. For example, in channels from $-$43~\kms{} to $-$35~\kms{} the jet appears as a chain of several jet knots. 

\autoref{fig:outflow_chan} also reveals some unusual properties of this jet. Firstly, 
instead of a continuous linear feature, the jet seems to be composed of a few segments with slightly different directions. This is most clear at panels from $-$35 to $-$19~\kms{}, and from 37 to 69~\kms. Secondly, at velocities from $-$11 to $-$3~\kms{} the jet gradually turns into a wide angle V-shape outflow with a half opening angle of $\sim$20\arcdeg. The coexistence of both a collimated jet-like component and a wide-angle biconical component has been observed in low-mass outflows \citep[e.g., IRAS 04166+2706, L1448C, HH 212,][]{Santiago-Garcia09, Hirano10, Codella14}. However, in these sources the jet is usually located at the central axis of the wide angle shell, while for IRS3 the jet is spatially offset from the axis of the wide angle component. This wide angle outflow is not apparent in the redshifted lobe at the corresponding velocity range, i.e., from 21 to 30~\kms.

\autoref{fig:outflow_chan_IRS1} presents a zoom-in view of the CO outflow associated with IRS1. The IRS1 outflow is more prominent in the blueshifted lobe from $-$23 to $-$3~\kms. It appears as V-shape centered on IRS1 at higher velocities (i.e., panel $-$23, $-$19~\kms), with its opening facing towards the SW direction, and turns more like a bubble in shape at lower velocities, which has a radius of $\sim$ 1\farcs{8}. Such a bubble-like feature is not seen in the redshifted lobe. At higher velocities, i.e., $-$39 to $-$27~\kms, the blueshifted outflow turns into a clump, about 0\farcs{6} to the west of IRS1. The redshifted lobe is more complex. There is some weak CO emission originating from IRS1 extending to the NE direction up to around 0\farcs{8} at velocities from 25 to 33~\kms. This redshifted CO emission may be driven by the same source that is responsible for the blueshifted bubble outflow given their roughly aligned direction, but it is unclear why they have such dramatically different appearances. At higher velocities there is a redshifted clump around 1\farcs{5} to the east of IRS1 from 25 to 57~\kms. While this clump appears as a seemingly continuous feature in velocity, detailed inspection suggests that it is actually composed of three clumps with narrower velocity ranges, and the clumps with higher velocities are located more to the NE direction of IRS1. The first clump R1 spans a velocity range from 25~\kms{} to 37~\kms. While R1 turns very weak at around 41~\kms, another CO clump (R2) emerges from 41 to 45~\kms, which is very close to, but slightly offset with R1. The highest velocity clump R3 is more prominent from 49 to 57~\kms{} and is clearly offset from R1 and R2. The positions of the three clumps are also marked on \autoref{fig:outflow_chan_IRS1} and \autoref{fig:outflow_sum}. 

It is worth noting that there is some blueshifted emission that is most apparent from $-$19~\kms{} to $-$3~\kms{} to the northeast of IRS1, as seen in \autoref{fig:outflow_chan}. This outflow component has a wide angle morphology and coincides well with the \htwo{} outflow IIA identified in \citet{Eisloffel00}, which is driven by IRS1. Thus this wide angle CO outflow could also be associated with IRS1, although the vertex of the wide angle cavity appears to be offset from IRS1 by $\sim$5\arcsec. These complicated outflow detections further reveal the complexity in the accretion and ejection process in the vicinity of the IRS1 protostar(s).

\autoref{fig:outflow_sum} provides an overview plot of the outflow detections for IRS1 and IRS3. The plot is overlaid on a three color image made with integrated blueshifted and redshifted CO emission, as well as the 1.3~mm continuum (in green). We have classified the detections into ``high-v'' and ``low-v'' and labeled them with different colors. This classification is based on the main velocity range of the outflow detections, using $| v - v_{\rm sys}|$ = 25~\kms{} as a dividing point. In summary, both IRS1 and IRS3 exhibit a variety of outflow morphologies at different velocities and there are indications of changes on the ejection direction for both sources. We further overlaid on \autoref{fig:outflow_sum} the directions of the radio jets inferred from this work or literature, which are shown in cyan lines. For IRS3 the direction of radio jet is not consistent with either the high velocity or low velocity CO outflow. For IRS1 the radio jet is close to the high-v outflow. The situation becomes more complicated with the inclusion of disk orientation inferred from continuum and/or line kinematics. While the IRS3 disk is broadly consistent with the radio/molecular ejections, for IRS1 the ejection direction inferred from disk orientation is offset from the observed radio/high-v outflow by $\gtrsim$ 50\arcdeg. We further discuss possible origins in \autoref{sec:discussion_outflow}.

\begin{figure*}[ht!]
\epsscale{1.1}\plotone{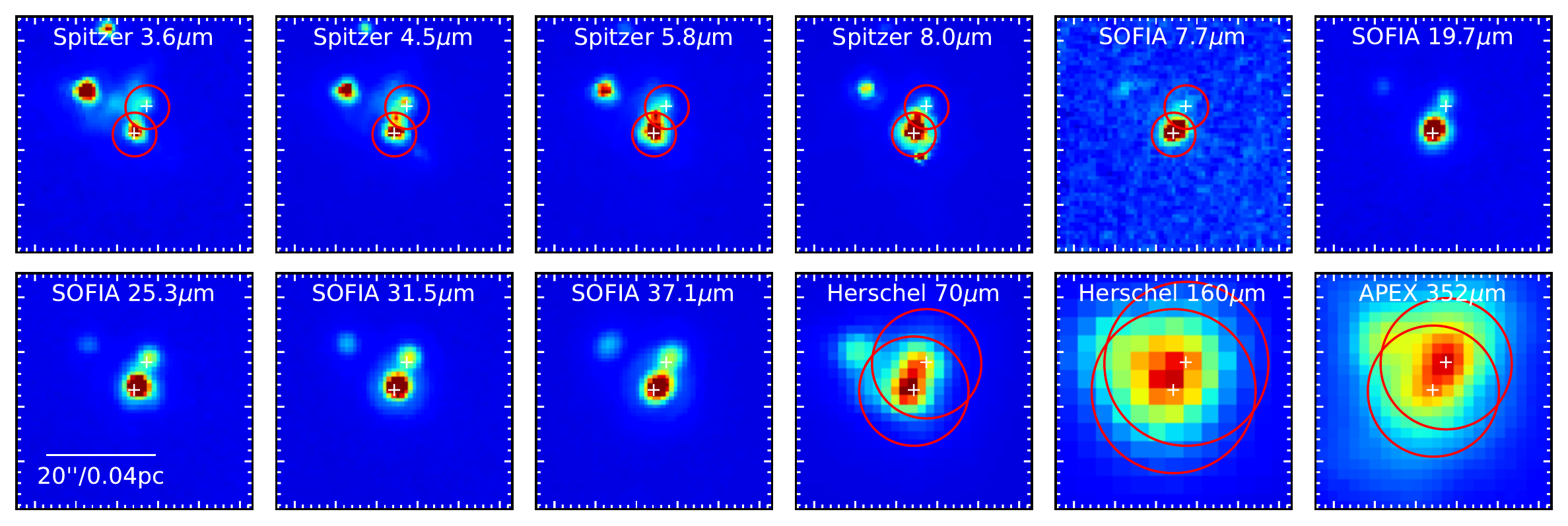}
\caption{Maps of IRS1 and IRS3 in different wavelengths observed with {\it Spitzer}, {\it SOFIA} and {\it Herschel}. The positions of IRS1 and IRS3 are marked with white crosses. The red circles indicate the aperture used for photometry. For the {\it SOFIA} 19.7~$\mu$m, 25.3~$\mu$m, 31.5~$\mu$m and 37.1~$\mu$m images we perform a 2D Gaussian fitting towards IRS1 and IRS3 to better measure their fluxes.\label{fig:multi_wave}
}
\end{figure*}

\begin{figure*}[ht!]
\epsscale{1.1}\plotone{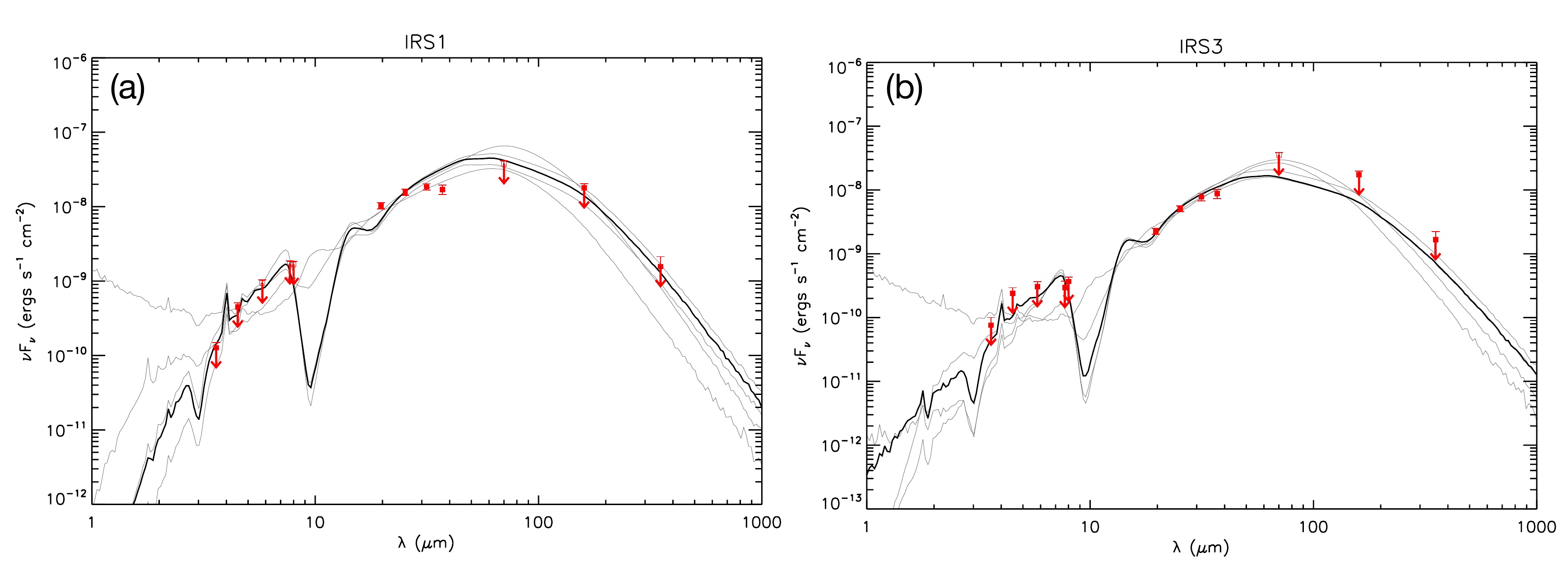}
\caption{Protostar model fitting to the fixed aperture, background-subtracted SED of IRS1 and IRS3 using the ZT model grid. The best-fit model is shown with a solid black line and the next four best models are shown with solid gray lines.\label{fig:sed}
}
\end{figure*}

\section{SED analysis}
\label{sec:sed}

To provide more constraints on the physical parameters of IRS1 and IRS3 like protostellar masses, we performed a SED fitting towards the two sources with data from near-IR to sub-mm band. Similar analysis has been conducted in \citet{Liu20} towards NGC~2071~IR but in their fiducial case a fixed aperture of 9\farcs{6} centered on IRS1 is adopted, which encompasses the emission of both IRS1 and IRS3. Here we follow the same fitting routine as in \citet{Liu20} but attempt to separate the flux between IRS1 and IRS3. We retrieved the same dataset, i.e., {\it Spitzer/IRAC} 3.5, 4.5, 7.3, 8.0~$\mu$m, {\it SOFIA/FORCAST} 7.7, 19.7, 25.3, 31.5, 37.1~$\mu$m and {\it Herschel/PACS} 70, 160~$\mu$m map as in \citet{Liu20} (see references therein). Additional {\it APEX/SABOCA} 352~$\mu$m data are also included here. In \autoref{fig:multi_wave} we present the multi-wavelength images of IRS1 and IRS3. In short wavelengths, i.e., 3.6 to 8.0~$\mu$m, IRS1 is clearly seen while the detection IRS3 is relatively weak. At mid-IR wavelengths from 19.7 to 37.1$\mu$m IRS1 is still the primary flux contributor and IRS3 is also apparent. At longer wavelengths, the resolution of \textit{Herschel} is not sufficient to resolve the two objects.

In order to better disentangle the flux emitted by IRS1 and IRS3, we performed the photometry in a ``heterogeneous'' way for data at different wavelengths. For wavelengths from 19.7 to 37.1~$\mu$m, where both objects are clearly detected and partly blended, we performed a two component 2D Gaussian fitting to obtain their fluxes. For shorter wavelengths we did aperture photometry with a 4\arcsec{} aperture, which is chosen to cover vast majority of emission from each object. Following the routine in \citet{Liu20}, we carry out a background subtraction using the median flux density in an annular region extending from one to two aperture radii, to remove general background and foreground contamination. Note that for IRS3, the flux measurement is most likely overestimated since the adopted aperture also covers part of the emission from IRS1, and there is some contamination of extended nebulosity in 3.6 and 4.5~$\mu$m. This will not significantly affect our SED fitting results since in our SED modeling the data points of shorter wavelength ($<$ 8.0~$\mu$m) are treated as upper limits. For longer wavelengths (70, 160, 352~$\mu$m), it is difficult to disentangle the fluxes of blended sources, i.e., IRS1 and IRS3 (and potentially IRS2) with the current resolution. In light of this, we performed an aperture photometry with an aperture that is large enough to encompass the fluxes of both IRS1 and IRS3 (12\arcsec{} in 70$\mu$m, 15\arcsec{} in 160$\mu$m, 10\arcsec{} in 352~$\mu$m) and set the data points as upper limits when performing the SED fitting. 

We use \citet{Zhang18} radiative transfer models (hereafter ZT models) to fit the SEDs and derive key physical parameters of the protostars. The ZT model is a continuum radiative transfer model that describes the evolution of high- and intermediate-mass protostars with analytic and semi-anlytic solutions based on the paradigm of the Turbulent Core model \citep[see][for more details]{Zhang18}. The main free parameters in this model are the initial mass of the core $M_c$, the mass surface density of the clump that the core is embedded in $\Sigma_{\rm cl}$, the protostellar mass $m_*$, as well as other parameters that characterize the observational setup, i.e., the viewing angle $i$, and the level of foreground extinction $A_V$. Properties of different components in a protostellar core, including the protostar, disk, infall envelope, outflow, and their evolution, are also derived self-consistently from given initial conditions. In \autoref{table:sed}, we present the parameters of five best-fit models, ordered from best to worst as measured by $\chi^2$. The best-fit SEDs are shown in \autoref{fig:sed}.

The best fitting model of IRS1 indicates a source with a protostellar mass of 4~\msun{} accreting at a rate of 3$\times10^{-5}$ \msunyr{} inside a core with an initial mass of 40~\msun{} embedded in clumps with a mass surface density of 0.1 \gcm. Nevertheless, the best five models provide similar goodness of fit, judging from the value of $\chi^2$, although there is a significant variation in model parameters like the protostellar mass $m_*$. For example, similar $\chi^2$ can be achieved with a protostellar source of mass $\sim$ 1~\msun{} accreting at 6$\times10^{-5}$ \msunyr. This illustrates the model degeneracy that exists in trying to constrain protostellar properties from only their MIR to FIR SEDs \citep[see also][]{Buizer17}. In light of this we only consider the typical parameter ranges among the best five models as a reasonable initial constraint for the protostellar system, instead of exploring only the best fitting case. Therefore, IRS1 can be fitted with a protostellar source with a central mass of 1 -- 4~\msun, with an accretion rate of 3 -- 19 $\times 10^{-5}$~\msunyr. Similarly, the SED of IRS3 is described with a protostellar source with a central mass of 1 -- 4~\msun, with an accretion rate of 2 -- 19 $\times 10^{-5}$~\msunyr. The viewing angle, ranging from 44\arcdeg{} to 71\arcdeg, is broadly consistent with the value derived from ALMA observations (67\arcdeg). Moreover, for IRS3 the best fit case provides a significantly better fitting ($\chi^2 \sim$ 0.32) compared with other four models; hence the returned parameters, including a protostellar mass with $m_*$ $\sim$ 2~\msun, is more favored. The relatively large uncertainty in the stellar mass inferred from the SED fitting highlights the need for an independent method for constraining this important quantity, e.g., through disk kinematics. 

A caveat in the SED analysis is that we have implicitly assumed the fluxes are mainly contributed by a single protostar, which may not be true for IRS1 or IRS3. IRS3 clearly contains a binary system, although the emission from the secondary component is much weaker in 9~mm. IRS1 appears single but may also host an unresolved multiple system, as hinted by the complicated outflow detections (see also \autoref{sec:discussion_outflow}). It is difficult to quantify how the resulting physical parameters vary if there are multiple components with comparable protostellar masses. One might expect a larger (total) stellar mass in this case to account for the same bolometric luminosity, given a typical luminosity-mass relation of $L\propto M_*^{4}$ for intermediate mass main sequence stars \citep[e.g.,][]{Eker15}. But this argument could be complicated by the protostellar luminosity evolution and the fact that a substantial fraction of luminosity may come from accretion.

\startlongtable
\begin{deluxetable*}{ccccccccccccc}
\tabletypesize{\scriptsize}
\renewcommand{\arraystretch}{1.0}
\tablecaption{Estimated physical parameters of IRS1 and IRS3 from SED fitting$^a$\label{table:sed}}
\tablehead{
\colhead{Source} & \colhead{$\chi^2$}  & \colhead{$M_c$} & \colhead{$\Sigma_{\rm cl}$} & \colhead{$m_*$} & \colhead{$i$} & \colhead{$A_V$} &  \colhead{$\theta_{\rm w,esc}$} & \colhead{$m_{\rm disk}^{b}$} & \colhead{$R_{\rm disk}$} &  \colhead{$ {\dot{m}_{\rm disk}}$} & \colhead{$L_{\rm bol,iso}$} & \colhead{$L_{\rm bol}$} \\
\colhead{} & \colhead{}  & \colhead{$M_\odot$} & \colhead{$\rm g cm^{-2}$} & \colhead{$M_\odot$} & \colhead{\arcdeg} & \colhead{mag} &  \colhead{\arcdeg} & \colhead{$M_\odot$} & \colhead{(au)} &  \colhead{\msunyr} & \colhead{$L_\odot$} & \colhead{$L_\odot$}
}
\startdata
IRS1 & 4.02 & 40 & 0.1 & 4.0 & 62 & 19.3 & 27 & 1.3 & 123 & 3.0 $\times10^{-5}$ & 0.8 $\times10^3$ & 0.4 $\times10^3$ \\
 & 4.62 & 10 & 1.0 & 1.0 & 29 & 16.8 & 25 & 0.3 & 19 & 6.0 $\times10^{-5}$ & 0.8 $\times10^3$ & 0.6 $\times10^3$ \\
 & 4.69 & 30 & 0.1 & 4.0 & 65 & 21.0 & 33 & 1.3 & 136 & 2.7 $\times10^{-5}$ & 0.8 $\times10^3$ & 0.4 $\times10^3$ \\
 & 5.27 & 10 & 3.2 & 4.0 & 62 & 0.0 & 56 & 1.3 & 39 & 19.0 $\times10^{-5}$ & 1.9 $\times10^3$ & 0.3 $\times10^3$ \\
 & 5.39 & 50 & 0.1 & 4.0 & 51 & 32.7 & 24 & 1.3 & 115 & 3.2 $\times10^{-5}$ & 0.8 $\times10^3$ & 0.5 $\times10^3$ \\ \hline
IRS3 & 0.32 & 30 & 0.1 & 2.0 & 58 & 6.7 & 23 & 0.7 & 79 & 2.0 $\times10^{-5}$ & 0.2 $\times10^3$ & 0.2 $\times10^3$ \\
 & 0.74 & 10 & 1.0 & 2.0 & 44 & 33.5 & 39 & 0.7 & 34 & 7.5 $\times10^{-5}$ & 0.8 $\times10^3$ & 0.3 $\times10^3$ \\
 & 0.76 & 40 & 0.1 & 2.0 & 55 & 16.8 & 19 & 0.7 & 73 & 2.2 $\times10^{-5}$ & 0.3 $\times10^3$ & 0.2 $\times10^3$ \\
 & 0.83 & 30 & 0.1 & 1.0 & 48 & 0.0 & 15 & 0.3 & 48 & 1.5 $\times10^{-5}$ & 0.2 $\times10^3$ & 0.1 $\times10^3$ \\
& 1.04 & 10 & 3.2 & 4.0 & 71 & 0.0 & 56 & 1.3 & 39 & 19.0 $\times10^{-5}$ & 1.9 $\times10^3$ & 0.2 $\times10^3$ \\
\enddata
 \tablenotetext{a}{From left to right, the parameters are reduced $\chi^2$, the initial core mass $M_c$, the mean mass surface density of the clump $\Sigma_{\rm cl}$, the current protostellar mass $m_*$, the viewing angle $i$, foreground extinction $A_V$, half opening angle of the outflow cavity $\theta_{\rm w,esc}$, the mass of the disk $m_{\rm disk}$,  the radius of the disk $R_{\rm disk}$, accretion rate from the disk to the protostar $\dot{m}_{\rm disk}$, the luminosity integrated from the unextincted model SEDs assuming isotropic radiation $L_{\rm bol,iso}$, and the inclination-corrected true bolometric luminosity $L_{\rm bol}$. }
 \tablenotetext{b}{In the ZT SED model, the ratio between disk mass and protostellar mass, $m_{\rm disk}/m_*$, is fixed to be 1/3.}
\end{deluxetable*}

\section{Kinematic Modeling}\label{sec:kine}

The detection of molecular lines has provided an opportunity to quantify the gas kinematics and more precisely measure the stellar masses. Different approaches have been developed to measure the dynamical mass based on the kinematic structure of molecular lines.
A widely adopted method uses PV diagrams to fit Keplerian rotation,
either the outer edge or the intensity maxima \citep[see][for a discussion]{Seifried16}. More sophisticated modelling that includes a proper parameterization of the physical structure of the disk, e.g., temperature and density profile, and radiative transfer with code like RADMC-3D \citep{Dullemond12}, has also been developed \citep[e.g.,][]{Czekala15,Czekala16,Sheehan19}. However, they can be computationally expensive
and may have difficulties in the presence of considerable extended emission and multiple sources like the case in NGC~2071~IR.
Similarly accurate determination of dynamical mass could be achieved via pure kinematic modeling without detailed treatment of the underlying physical structure of the disk \citep{Boyden20}. Here we develop a simple analytic model, which is similar as the one in \citet{Boyden20}, to interpret the observed PV diagrams and to infer the dynamical masses. 

\subsection{A simplified analytic model}
\label{sec:analytic_model}

The observed kinematics of molecular lines could arise from two components: a Keplerian rotating disk, and an infalling-rotating envelope. The envelope is further discussed in \autoref{sec:mcmc_env}. For the disk, we assume an optically thin, uniformly excited disk orbiting around a central object with mass $m_*$ and we assume that the disk has a height $h(r)$ = 0.2$\times r$ on both sides of the midplane, and a sharp truncation at the inner boundary $R_{\rm in}$, and an outer radius $R_{\rm disk}$. The density distribution of the disk is described by $\rho \propto r^{-2.5}$. This corresponds to a surface density $\Sigma(r) \propto r^{-1.5}$, i.e., similar as those found in protoplanetary disks \citep{Sheehan19}. The disk follows Keplerian rotation, i.e,

\begin{equation}
  v_r = 0,
\end{equation}
\begin{equation}
  v_\phi = \sqrt{\frac{Gm_*}{r}}.
\end{equation}


Based on the density and kinematic distribution above we can generate 3-D model grids to simulate the disk system, with each grid cell with a density $\rho_i$ and velocity $v_i$. To compare with observations we assume the disk is viewed at an inclination angle $i$ ($i$ = 0\arcdeg{} corresponds to a face-on configuration while $i$ = 90\arcdeg{} is edge-on). Thus we may calculate the line of sight velocity $v_{\rm i,los}$ for each grid cell depending on the viewing angle and the position of the model grid. Each grid then produces line emission with a velocity profile described by
\begin{equation}
\phi_{\rm v,obs} \propto \exp \left[-\frac{(v_{\rm obs} - v_{\rm i,los})^2}{2\sigma^2}\right],
\end{equation}
where we adopt $\sigma$ = $\sqrt{kT/\mu m_H}$, i.e., the thermal broadening line width, where $\mu$ is the molecular weight of the molecule in use. For temperature we use the dust temperature estimated in \autoref{sec:cont}. For a specific sky location, the line intensity at $v_{\rm obs}$ can be obtained by integrating $\phi_{\rm v,obs}\cdot \rho_i$ for each model grid along line of sight. In this way we can generate a position-position-velocity (PPV) cube. To mimic the observational setups we match the channel width of our model PPV cube to values in the observation (depending on which line is being fit), and smooth the PPV cube to the same spatial resolution as in the real observations. Then we extract a PV diagram along the disk midplane to compare with our observational results. Alternatively, we can also directly compare the model PPV cube with observations. In most cases these two methods give consistent results. Here we use the PV diagram, mainly because a PPV cube 
may also contain line emission from other structures other than the disk, especially for IRS1 (see \autoref{fig:lines_IRS1}). The PV diagram is extracted following the center position and orientation defined by the 1.3~mm continuum from 2D Gaussian fit (for IRS1 we follow a P.A.$\sim$ 135\arcdeg{} following the discussion in \autoref{sec:lines}). To match observations we include two additional parameters, $v_{\rm sys}$ and $\Delta x$, to describe the center of a PV diagram. For example, a non-zero $\Delta x$ means the center of the PV diagram deviates from the reference point, i.e., in this case, the source position defined by 2D Gaussian fit of the 1.3~mm continuum. 

\begin{figure*}[ht!]
\epsscale{1.1}\plotone{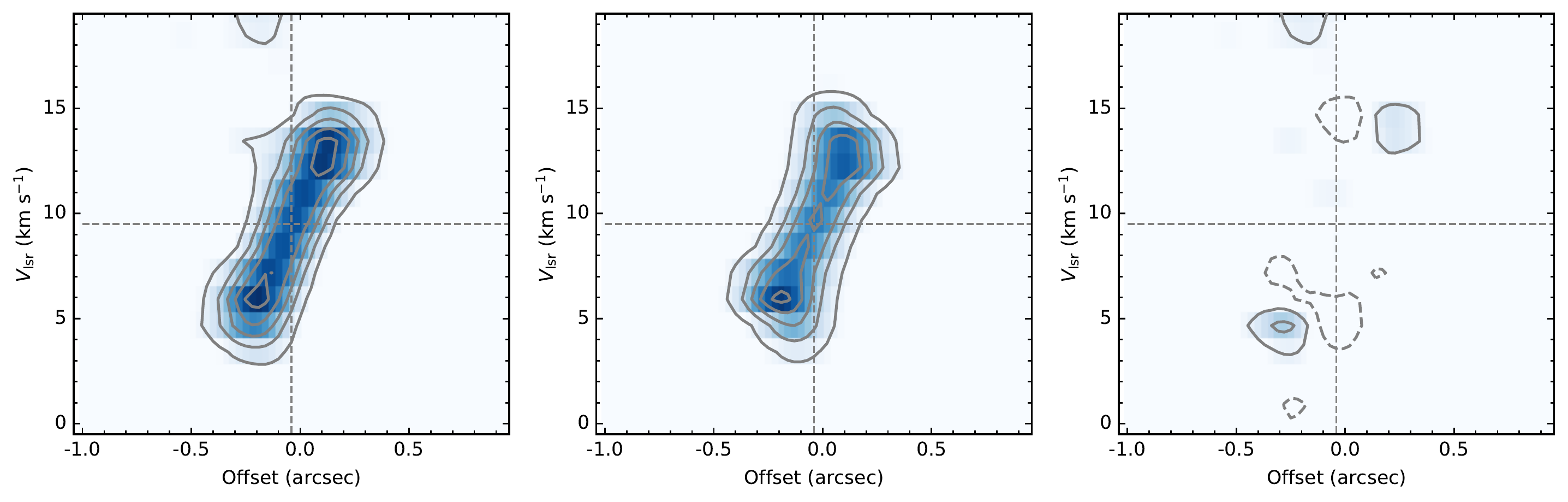}
\caption{ {\it Left:} PV diagram of $\rm CH_3OCHO ~ 18_{4,14}-17_{4,13}$ for IRS3 shown in colorscale and contours. {\it Middle:} The analytic model for a Keplerian rotating disk with $m_*$=1.5~\msun, $R_{\rm disk}$ = 130~au, $R_{\rm in}$ = 25~au, $v_{\rm sys}$ = 9.5~\kms{} and $\Delta x$ = -17~au. {\it Right:} The residual between model and observed PV diagrams. In all three panels the intensity is normalized by the peak intensity in observations, and the contours levels are (0.1, 0.3, 0.5, 0.7, 0.9). The horizontal and vertical dashed lines indicates the systemic velocity $v_{\rm sys}$ and position offset $\Delta x$, respectively. 
\label{fig:model}. 
}
\end{figure*}

In \autoref{fig:model} we present an example of the model output and compare it with the PV diagram of IRS3 from line $\rm CH_3OCHO ~ 18_{4,14}-17_{4,13}$, which is selected as a representative for group B lines described in \autoref{sec:lines}. The model assumes a Keplerian rotating disk with $m_*$ = 1.5~\msun, $R_{\rm disk}$ = 130~au, $R_{\rm in}$ = 25~au and $v_{\rm sys}$ = 9.5~\kms. This model appears reasonably consistent with emission of $\rm CH_3OCHO ~ 18_{4,14}-17_{4,13}$, suggesting that it indeed can be explained by a Keplerian rotating disk. Such disk-only models may not fit the group A lines well as they show significant extended low velocity emission and may have a contribution from the inner envelope, which is further discussed in \autoref{sec:mcmc_env}

\subsection{Dynamical mass estimation}
\label{sec:mass_estimation}

For the purpose of dynamical mass estimation we utilize the group B lines, i.e., lines that unambiguously trace the disk. This allows us to reduce the number of free parameters, and also the systematic uncertainties since the disk kinematics are much simpler. To further assess the fit quality we use a $\chi^2$ likelihood, defined as

\begin{equation}
\chi^2 = \sum \left( \frac{{\rm Data}(x,v)-{\rm Model}(x,v)}{\sigma} \right)^2,
\end{equation}
i.e., the sum of $\chi^2$ over all pixels within a localized region in the PV diagram. $\sigma$ is the rms noise of the PV diagram measured with signal-free regions. 

In summary, we have in total seven parameters for the disk model, \{$m_*$, $R_{\rm disk}$, $R_{\rm in}$, $i$, $v_{\rm sys}$, $\Delta x$, $f_{\rm norm}$\}, including the stellar mass $m_*$, disk inner/outer radius $R_{\rm in}$ and $R_{\rm disk}$, inclination $i$, systemic velocity $v_{\rm sys}$ and position offset $\Delta x$. Since the model does not provide any constraints on the absolute intensity we also include a normalization factor $f_{\rm norm}$ to compare with observations. For convenience in the calculations both model and observed PV diagrams are normalized by their peak intensities, so the $f_{\rm norm}$ factor is $\sim$1. In order to explore the parameter space more effectively we adopt the Markov Chain Monte Carlo (MCMC) fitting code {\it emcee} \citep{Foreman-Mackey13}. Uniform prior probability distributions for the parameters is assumed, for the mass $m_*$ in a range of 1 -- 15 \msun, $R_{\rm in}$ in a range of 0 -- 40~au, $R_{\rm disk}$ in a range of 40--200~au, $i$ in a range of 0 -- 90 degrees, $v_{\rm sys}$ in a range of 8 -- 12 \kms{}, $\Delta x$ in a range of $-$40 -- 40~au and $f_{\rm norm}$ in a range of 0.7 -- 1.3.




We selected lines that are relatively strong and isolated so that we can safely avoid the contamination from other lines that are close in frequency. These are $\rm CH_3OH ~ 18_{3,15}-17_{4,14}$, $\rm CH_3OH ~ 10_{3,7}-11_{2,9}$, $\rm ^{13}CH_3OH ~ 5_{1,5}-4_{1,4}$, $\rm SO_2~ 28_{3,25}-28_{2,26}$ and $\rm CH_3OCHO ~ 18_{4,14}-17_{4,13}$. We run the MCMC routine for the PV diagram of each line separately. In practice we found that there is usually some coupling between $m_*$ and $i$, which is expected. Consider a narrow ring with radius $r$ rotating around a center mass $m_*$, then on the PV diagram one would get a velocity gradient

\begin{equation}
\frac{\partial v_{\rm LOS}}{\partial x} = \sqrt{\frac{Gm_*}{r^3}} \sin (i).
\end{equation}
The case for an inner truncated disk is similar, i.e., equivalent to a set of rings with radii from $R_{\rm in}$ to $R_{\rm disk}$. So our model is not very effective at optimizing both $m_*$ and $i$ simultaneously, especially when the disk kinematics are not well-resolved. In this case we may overfit the data and the walkers could struggle to achieve global optimization when multiple $\chi^2$ minimums exist. 
Therefore, we attempted two strategies of parameter setup: one with the inclination $i$ as a free parameter and a prior range of 0 -- 90\arcdeg{} is given. As for the second strategy, we use a fixed $i$ to avoid overfitting. For IRS3 we adopt the inclination angle inferred from the dust continuum (i.e., 67\arcdeg). For IRS1 the inclination is not well constrained so we fixed the inclination angle for a range of discrete values, i.e., 90\arcdeg, 60\arcdeg{} and 30\arcdeg, and then run MCMC routine.

\autoref{table:mcmc} lists the best-fit parameters for IRS1 and IRS3. We found that different lines return broadly consistent disk parameters. In the fixed-$i$ case, our modeling of IRS3 gives a mass ranging from 1.43 to 1.52~\msun, and a radius ranging from 112 to 134~au for the different lines. This mass estimation is in reasonable agreement with the SED model results for IRS3, i.e, $\sim$ 2.0~\msun. The radius estimation also agrees well with the value derived from millimeter continuum ($\sim$103~au). $\rm CH_3OH ~ 18_{3,15}-17_{4,14}$ gives the lowest mass value of 1.43 $\pm$ 0.02~\msun{}, while $\rm CH_3OH ~ 10_{3,7}-11_{2,9}$ gives the highest mass value of 1.52 $\pm$ 0.18~\msun. The models with free inclination work reasonably well and give very similar results as models with fixed $i$. The best-fit $i$ ranges from 63 to 77\arcdeg, indicating that a configuration that is close to edge-on is also favored from the modeling. These values appear in good agreement with the fiducial inclination of 67\arcdeg.

For IRS1 the resultant mass strongly depends on the assumed inclination value. For a moderate inclination $i$ = 60\arcdeg, the modeling returns a mass of 4.40 -- 5.46~\msun{} among different lines. High inclination (90\arcdeg) results in a lower mass estimation, i.e, 3.68 -- 4.30~\msun, while low inclination $i$=30\arcdeg{} tends to give a high mass estimation around 10.03 -- 14.06~\msun, which is unrealistically large for IRS1 given the observed bolometric luminosity. So lower inclinations (i.e., more close to face-on configuration) are not explored. The free-$i$ models further put some constraints on the inclination, with the best-fit values ranging from 56\arcdeg{} to 70\arcdeg{} for different lines. The corresponding masses span from 3.58 to 5.65~\msun. We discuss possible range of stellar masses of IRS1 in \autoref{sec:mass}.

\startlongtable
\begin{deluxetable*}{llccccccc}
\tabletypesize{\scriptsize}
\renewcommand{\arraystretch}{1.0}
\tablecaption{Estimated physical parameters of IRS3 and IRS1 from the kinematic modeling\label{table:mcmc}}
\tablehead{
\colhead{Source} & \colhead{Molecule/Transition} & \colhead{$i$} &\colhead{$m_*$} & \colhead{$R_{\rm disk}$} & \colhead{$R_{\rm in}$} &\colhead{$ v_{\rm sys}$}  & \colhead{$\Delta x$} & \colhead{$f_{\rm norm}$} \\
\colhead{} & \colhead{} &\colhead{(degree)} & \colhead{($M_\odot$)} & \colhead{(au)} & \colhead{(au)} & \colhead{\kms} & \colhead{(au)} &\colhead{}
}
\startdata
IRS3&	$\rm CH_3OH ~ 18_{3,15}-17_{4,14}$&	72.54$\pm$7.41&	1.38$\pm$0.09&	118.53$\pm$3.40&	14.07$\pm$0.99&	9.40$\pm$0.04&	-14.46$\pm$1.33&	1.14$\pm$0.03\\
&		&67&	1.43$\pm$0.02&	120.86$\pm$3.01&	13.69$\pm$1.05&	9.42$\pm$0.04&	-14.55$\pm$1.14&	1.13$\pm$0.03\\
&	$\rm CH_3OH ~ 10_{3,7}-11_{2,9}$&	67.95$\pm$6.70&	1.47$\pm$0.11&	129.59$\pm$5.01&	10.98$\pm$2.71&	9.43$\pm$0.11&	-19.45$\pm$1.44&	1.11$\pm$0.04\\
&		&67&	1.52$\pm$0.18&	133.85$\pm$4.39&	14.20$\pm$2.56&	9.47$\pm$0.10&	-19.54$\pm$1.81&	1.08$\pm$0.05\\
&	$\rm ^{13}CH_3OH ~ 5_{1,5}-4_{1,4}$&	76.73$\pm$8.49&	1.53$\pm$0.16&	136.01$\pm$9.12&	18.89$\pm$0.71&	9.34$\pm$0.13&	-13.51$\pm$2.00&	1.09$\pm$0.04\\
&		&67&	1.49$\pm$0.13&	129.72$\pm$7.19&	16.53$\pm$1.87&	9.41$\pm$0.04&	-13.60$\pm$1.16&	1.16$\pm$0.03\\
&	$\rm SO_2~ 28_{3,25}-28_{2,26}$ &	63.46$\pm$3.45&	1.57$\pm$0.08&	110.96$\pm$12.68&	10.42$\pm$1.29&	9.49$\pm$0.04&	-16.96$\pm$2.07&	1.18$\pm$0.05\\
&		&67&	1.49$\pm$0.01&	112.42$\pm$4.58&	10.42$\pm$0.73&	9.47$\pm$0.02&	-18.25$\pm$1.36&	1.24$\pm$0.04\\
&	$\rm CH_3OCHO ~ 18_{4,14}-17_{4,13} $&	68.77$\pm$4.67&	1.50$\pm$0.03&	130.41$\pm$4.84&	25.87$\pm$1.92&	9.50$\pm$0.03&	-16.48$\pm$1.44&	1.14$\pm$0.03\\
&		&67&	1.50$\pm$0.02&	127.05$\pm$3.08&	25.48$\pm$0.52&	9.50$\pm$0.02&	-17.04$\pm$1.23&	1.16$\pm$0.03\\ \hline
IRS1&	$\rm CH_3OH ~ 18_{3,15}-17_{4,14}$&	55.63$\pm$4.82&	5.13$\pm$0.43&	139.75$\pm$7.30&	37.06$\pm$5.21&	8.70$\pm$0.06&	4.56$\pm$1.48&	1.18$\pm$0.06\\
&		&90&	3.76$\pm$0.08&	128.04$\pm$4.71&	43.64$\pm$2.52&	8.76$\pm$0.07&	2.88$\pm$1.36&	1.18$\pm$0.04\\
&		&60&	4.62$\pm$0.19&	138.63$\pm$6.31&	34.99$\pm$4.22&	8.76$\pm$0.07&	4.78$\pm$1.83&	1.15$\pm$0.04\\
&		&30&	12.24$\pm$0.26&	143.67$\pm$5.83&	33.53$\pm$1.72&	8.65$\pm$0.04&	4.48$\pm$1.38&	1.26$\pm$0.03\\
&	$\rm CH_3OH ~ 10_{3,7}-11_{2,9}$&	57.23$\pm$4.43&	4.73$\pm$0.41&	150.64$\pm$6.03&	40.80$\pm$1.79&	9.03$\pm$0.07&	-6.63$\pm$1.25&	1.07$\pm$0.03\\
&		&90&	3.82$\pm$0.13&	130.93$\pm$3.03&	54.19$\pm$2.07&	8.86$\pm$0.05&	-9.12$\pm$1.70&	1.08$\pm$0.04\\
&		&60&	4.58$\pm$0.36&	149.18$\pm$16.27&	42.78$\pm$16.53&	9.02$\pm$0.17&	-7.45$\pm$1.98&	1.07$\pm$0.06\\
&		&30&	11.87$\pm$0.26&	148.06$\pm$3.70&	38.91$\pm$0.93&	8.92$\pm$0.04&	-6.03$\pm$1.46&	1.16$\pm$0.03\\
&	$\rm ^{13}CH_3OH ~ 5_{1,5}-4_{1,4}$&	59.66$\pm$2.42&	5.49$\pm$0.23&	147.02$\pm$3.25&	59.91$\pm$1.29&	8.77$\pm$0.04&	-5.34$\pm$1.55&	1.17$\pm$0.03\\
&		&90&	4.30$\pm$0.11&	143.88$\pm$3.01&	63.23$\pm$1.72&	8.81$\pm$0.03&	-6.24$\pm$1.38&	1.04$\pm$0.03\\
&		&60&	5.46$\pm$0.09&	146.90$\pm$3.03&	59.70$\pm$1.40&	8.77$\pm$0.04&	-5.29$\pm$1.44&	1.17$\pm$0.03\\
&		&30&	13.73$\pm$0.19&	159.08$\pm$4.39&	48.20$\pm$1.36&	8.91$\pm$0.04&	-3.27$\pm$1.61&	1.19$\pm$0.03\\
&	$\rm SO_2~ 28_{3,25}-28_{2,26}$ &	69.92$\pm$4.09&	3.58$\pm$0.24&	130.76$\pm$7.32&	28.32$\pm$3.08&	9.35$\pm$0.12&	-8.22$\pm$1.66&	1.14$\pm$0.03\\
&		&90&	3.68$\pm$0.32&	143.50$\pm$7.96&	28.75$\pm$4.35&	9.13$\pm$0.22&	-8.39$\pm$1.66&	1.06$\pm$0.03\\
&		&60&	4.40$\pm$0.31&	153.82$\pm$7.38&	18.21$\pm$7.02&	9.14$\pm$0.15&	-6.03$\pm$1.36&	1.11$\pm$0.04\\
&		&30&	10.03$\pm$0.44&	157.70$\pm$13.23&	15.62$\pm$4.43&	9.47$\pm$0.09&	-6.37$\pm$1.46&	1.19$\pm$0.02\\
&	$\rm CH_3OCHO ~ 18_{4,14}-17_{4,13} $&	55.61$\pm$3.01&	5.65$\pm$0.37&	133.29$\pm$3.34&	56.90$\pm$2.15&	8.83$\pm$0.03&	-6.89$\pm$1.48&	1.19$\pm$0.04\\
&		&90&	4.10$\pm$0.17&	132.39$\pm$20.19&	63.40$\pm$16.48&	8.87$\pm$0.14&	-6.97$\pm$2.24&	1.00$\pm$0.04\\
&		&60&	5.34$\pm$1.94&	134.50$\pm$21.74&	59.40$\pm$12.33&	8.83$\pm$0.05&	-7.49$\pm$2.22&	1.14$\pm$0.05\\
&		&30&	14.06$\pm$0.37&	129.42$\pm$6.24&	51.43$\pm$3.06&	8.83$\pm$0.14&	-5.68$\pm$1.61& 	1.26$\pm$0.05\\
\enddata
\end{deluxetable*}

\subsection{Kinematic modeling with both disk and envelope components}
\label{sec:mcmc_env}

Instead of a pure Keplerian rotating disk, the group A lines, i.e., those from
\ceighteeno, \thirteenco, \htwoco{} and \so{} are likely to have a contribution from
a surrounding envelope. To test this we apply our kinematic model to
group A lines by adding an envelope component. In this section we focus on the SO $\rm 6_5-5_4$ line since it has the most symmetric appearance among group A
lines (see the PV diagram in \autoref{fig:pv_IRS3}).
SO $\rm 6_5-5_4$ is also less affected by the optical depth issue compared to \ceighteeno~2--1 or \thirteenco~2-1. Only IRS3 is investigated here since it has a more symmetric disk appearance and we are able to independently estimate its inclination.

\begin{figure*}[ht!]
\epsscale{1.}\plotone{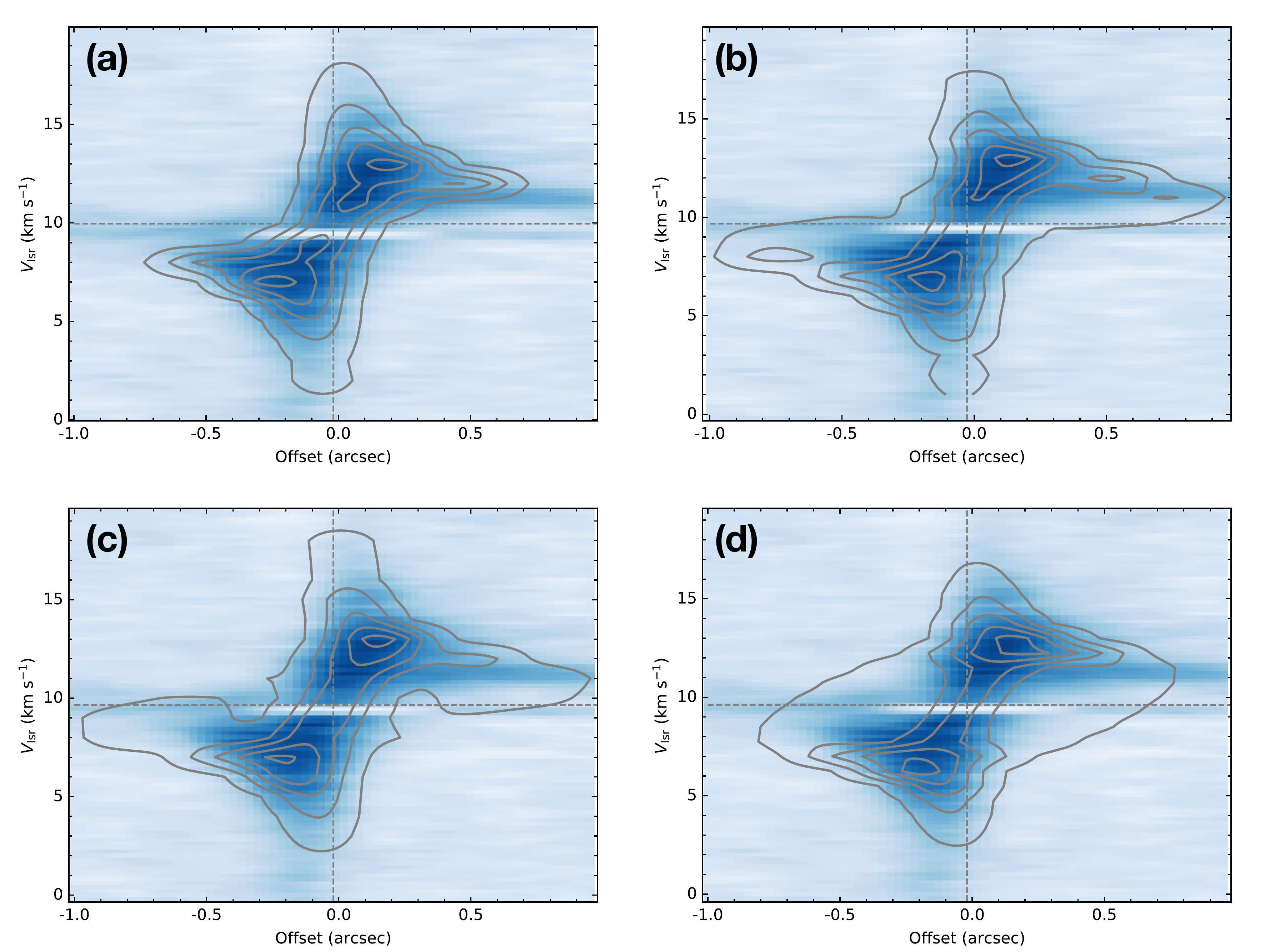}
\caption{PV diagram of SO $\rm 6_5-5_4$ of IRS3 overlaid with the predictions of the analytic model. {\it (a)} The contours indicate the analytic model for a Keplerian rotating disk. {\it (b)} The contours indicate the analytic model for a Keplerian rotating disk plus a simplified thin envelope model. {\it (c)} Same as {\it (b)} but use a modified rotation velocity profile for the envelope component, see text for more details. {it (d)} The contours indicate the analytic model for a Keplerian rotating disk plus an envelope with kinematic and density properties set by the ballistic solutions in \citet{Ulrich76}. In all panels the model is normalized by the peak intensity and the contour levels are (0.1, 0.3, 0.5, 0.7, 0.9). The parameters for the disk model in {\it (a)} are \{$m_*$, $R_{\rm disk}$, $R_{\rm in}$, $i$, $v_{\rm sys}$, $\Delta x$, $f_{\rm norm}$\}, i.e., the stellar mass $m_*$, disk inner/outer radius $R_{\rm in}$ and $R_{\rm disk}$, inclination $i$, systemic velocity $v_{\rm sys}$ and position offset $\Delta x$. For the disk$+$envelope model in {\it (b)}, {\it (c)}, and {\it (d)}, an parameter of $R_{\rm out}$ is introduced to characterize the outer boundary of the envelope. For the diks$+$Ulrich76 envelope model in {\it (d)} we have an additional parameter $m_{\rm env}/m_{\rm disk}$ to describe the mass ratio between disk and envelope (see text for more details). }\label{fig:model_env}
\end{figure*}

In \autoref{fig:model_env} and \autoref{table:mcmc_env} we present the best fit models for the PV diagram of SO $\rm 6_5-5_4$ line towards IRS3. First, to test if the PV diagram can be well
modeled by a pure disk we run the same MCMC routine as in \autoref{sec:mass_estimation}
with a fixed inclination angle of 67\arcdeg. We also allow for a larger range to search for
the disk radius $R_{\rm disk}$, i.e., a uniform prior probability distributions from 0 to 400~au.
The best fit model returns a stellar mass of 1.61~\msun{} and a radius of 269~au.
While the stellar mass is slightly larger, but still broadly consistent with the measurements in \autoref{sec:mass_estimation}, the disk radius is significantly larger, which is expected since the SO emission is more extended. The model fails to reproduce the low velocity emission extending beyond $\sim$0\farcs{5}, especially for the redshifted part.

\begin{figure}[ht!]
\epsscale{1.}\plotone{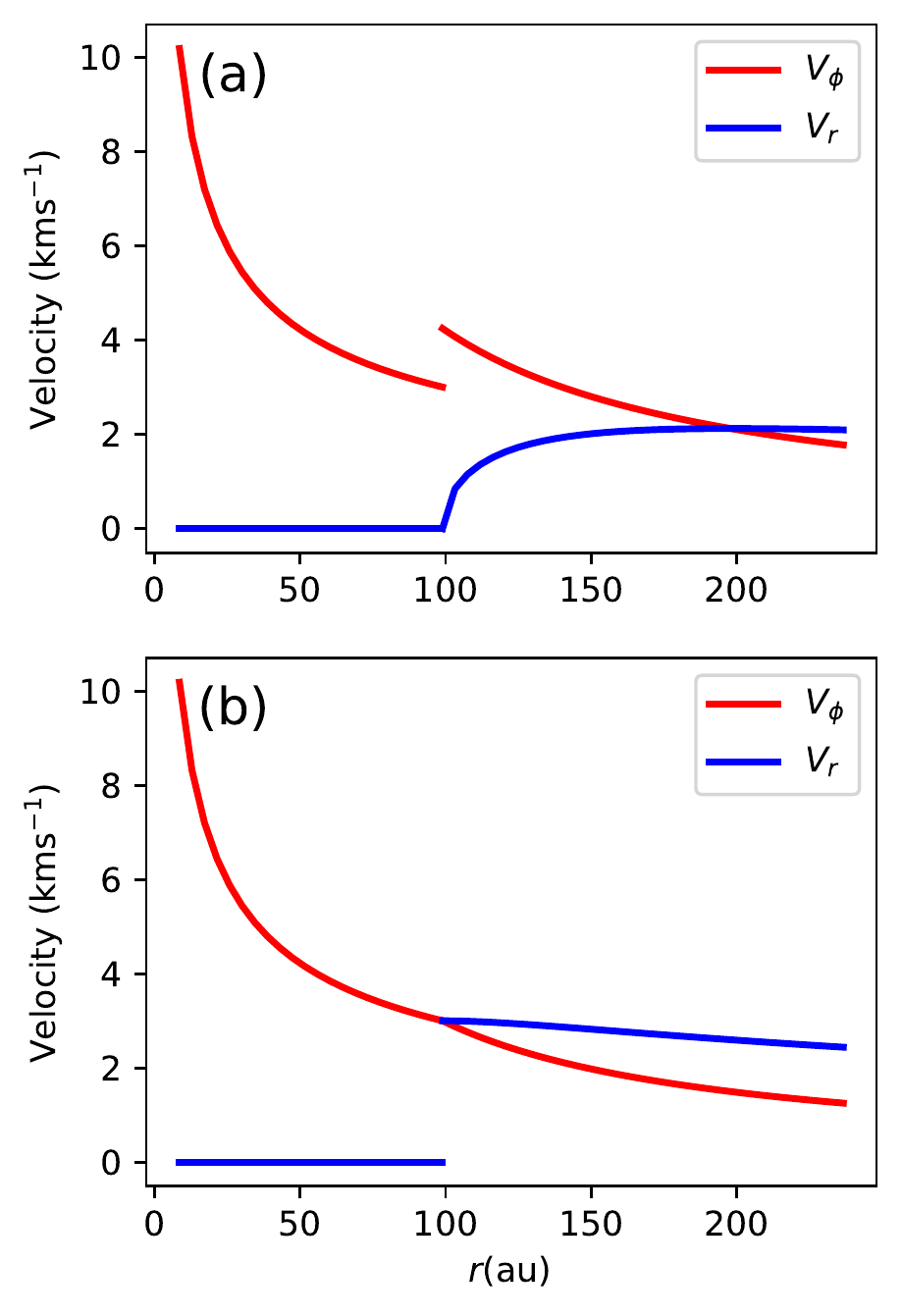}
\caption{An example of the velocity profiles adopted in the analytic model. Here we assume a stellar mass of 1~\msun{} and an edge-on disk with a radius of 100~au. {\it (a)} The envelope kinematics described in \autoref{sec:analytic_model}, i.e., the infalling and rotating material in the envelope reaches the centrifugal barrier ($R_{\rm disk}$ = $R_{\rm cb}$ = 100~au), where a disk forms. {\it (b)} A modified version of {\it (a)}. The radial and and tangential velocities are also described by \autoref{equ:vr} and \autoref{equ:vphi} but we assume Keplerian rotation motion inside the centrifugal radius ($R_{\rm disk}$ = $R_c$ = 2$R_{\rm cb}$ = 100~au). In this case the tangential velocity of the envelope will not exceed the Keplerian rotation velocity as in {\it (a)} and is continuous at the disk radius $R_{\rm disk}$.
\label{fig:vel}}
\end{figure}

\subsubsection{the simplified thin envelope model}

In light of this we add an envelope component starting from $R_{\rm disk}$ and extends to an outer
boundary $R_{\rm out}$. 
We assume the envelope has a flat geometry that is similar to the disk, i.e., density $\rho$ = 0 for $h$ > 0.2$\times$r.  The envelope starts from the centrifugal barrier at radius $R_{\rm disk}$ = $R_{\rm cb}$, and extends to an outer boundary $R_{\rm out}$. We also assume the density changes smoothly from the disk to envelope and the envelope follows a density distribution a $\rho \propto r^{-1.5}$. This corresponds to the typical density profile of an infalling cloud \citep[e.g.,][]{Shu77,Harvey03}. The envelope has the motion described by

\begin{equation}
  v_r = -v_{\rm cb}\frac{\sqrt{R_{\rm cb}(r-R_{\rm cb})}}{r},\label{equ:vr}
\end{equation}
\begin{equation}
  v_\phi = v_{\rm cb}\frac{R_{\rm cb}}{r},\label{equ:vphi}
\end{equation}

where $v_{\rm cb}$ = $\sqrt{2Gm_*/R_{\rm cb}}$ is the the rotation velocity at the centrifugal barrier. Such motion conserves both angular momentum and mechanical energy \citep[see, e.g.,][]{Sakai14}. An example of the adopted velocity profiles is shown in \autoref{fig:vel}. 

The same MCMC routine is run (with one more free parameter $R_{\rm out}$). The resultant best fit parameters are shown in \autoref{table:mcmc_env} (disk + envelope (A)). The PV diagram of the best fit model, which gives $m_*$ = 1.24~\msun, $R_{\rm disk}$ = 113~au and $R_{\rm out}$= 394~au, is shown in panel (b) of \autoref{fig:model_env}. This model has a better
performance in fitting the low velocity emission, and the best fit disk radius $R_{\rm disk}$
also agrees well with the measurements in \autoref{sec:mass_estimation}. This agrees with the explanation that group A lines (at least for SO) are tracing both the disk and the inner envelope components.

\subsubsection{the modified envelope model}
Interestingly, when an envelope component is added, the best fit stellar mass, 1.24$\pm$0.04~\msun{} is 
smaller than the best fit masses from modeling group B lines, i.e., 1.4 -- 1.5~\msun. 
This underestimation of dynamical
mass could be partly attributed to the oversimplified envelope kinematics in the calculation. For example,
the presented envelope model has ignored the motions of the infalling material in the $z$ direction.
In particular, the tangential velocity profile is not continuous at the location of $R_{\rm disk}$ (the radius
of centrifugal barrier), i.e., there is a jump by a factor of $\sqrt{2}$ when transitioning from the
disk to the envelope. Such velocity jump may not be physically realistic and has not been suggested in the observations.
Mathematically this discontinuity in the velocity profile could have led to the underestimation in central mass, since a smaller mass is needed to compensate for the large tangential velocity beyond $R_{\rm disk}$. 

To test this we run the MCMC routine with a modified velocity profile, in which we assume the envelope to disk transition takes place at the centrifugal radius, i.e., $R_{\rm disk}$ = $R_c$ = 2$R_{\rm cb}$ to avoid the rotation velocity keeps increasing inside $R_c$ and exceeding the Keplerian velocity. In this way the transition of rotation velocity profile becomes continuous (see \autoref{fig:vel}). The best fit results and parameters are presented in
\autoref{fig:model_env} and \autoref{table:mcmc_env} (disk + envelope (B)). As can be seen, 
this model returns $m_*$ = 1.44$\pm$0.09~\msun, consistent with the measurements using compact disk tracers in \autoref{sec:mass_estimation}.


\subsubsection{the Ulrich76 envelope model}
In order to further explore how the fitting is affected by different envelope models,
we also try to fit the observations with \citet{Ulrich76} (hereafter Ulrich76) envelope model. It is based on the solution
of the collapse of a spherically symmetric cloud in uniform solid-body rotation when
the pressure forces are negligible, in which material infalls following ballistic trajectories.
With the assumption of solid-body rotation, particles falling near the rotational axis have
smaller angular momentum and will fall in close to the central star, while particles falling in
from regions near $\theta \sim \pi/2$ will fall in to a maximum centrifugal radius $R_c$,
which is determined by the specific angular momentum measured around the rotation axis.
The material arriving at the midplane will collide with material arriving at the same position
from opposite $z$ direcion, thus producing a flat disk (after the kinematic energy dissipates in shocks).
For simplicity we ignore the detailed processes of disk formation and assume a thin disk
undergoing pure Keplerian rotating motion inside $R_c$. Therefore, for $r$ < $R_{\rm disk}$ (=$R_c$), we keep
the same geometry ($h$ = 0.2$\times r$), density and kinematic setup for the disk component as in
other models, and for  $r$ > $R_{\rm disk}$,  we adopt the ballistic solution as in Ulrich76 to set
the density and velocity for each model grid (the density is evaluated by assuming that
the mass infall rate is steady). However, there is a singularity in the density solution at $r$ = $R_c$
and $z$ = 0 (on the midplane). This will not cause numerical issues as long as no model grid is centered
on this point but we cannot assume a continuous density transition from the disk
to envelope as other models above. Instead we add another parameter, the mass ratio between the envelope and disk
$m_{\rm env}/m_{\rm disk}$, to properly characterize the density contrast between the two components.

The results are also shown in \autoref{fig:model_env} and \autoref{table:mcmc_env}.
The best fit model gives $m_*$ = 1.43~\msun, $R_{\rm disk}$ = 153~au and $R_{\rm out}$ = 347~au. Different
from the envelope models above, which shows a concave PV feature (i.e., at the second/fourth
quadrant in the PV diagram), the Ulrich76 envelope model seems to predict a more compact and diamond-shaped
PV diagram, in contrast with the observation. This appearance could partly arise from the difference in assumed
envelope geometry: in the simplified envelope case we have assumed a flat geometry
(i.e., density $\rho$ = 0 for $h$ > 0.2$\times$r), which is consistent with a relatively late evolutionary
stage when the surrounding material has been partly evacuated; in the Ulrich76 model there is still material distributed in smaller polar angles,
so the infalling material close to the rotational axis could contribute more low velocity emission, seen in the second/fourth quadrant of the PV diagram.

In summary, we attempt to model the SO $\rm 6_5-5_4$ line with different setups, including a pure Keplerian disk and a disk plus an inner envelope, and different envelope models are also investigated. In particular, the disk + simple envelope (A) model, in which we adopt the radius of centrifugal barrier as the transition point of the disk and envelope, appears to lead to an underestimation of the central stellar mass, mainly due to a jump in the rotation velocity profile. This suggests that the simplified thin envelope model based on the centrifugal barrier assumption may be over simplified in describing the kinematic transition between the disk and envelope.

Overall the model will have better performance in reproducing the low velocity extended emission when an envelope component is included, though the mass estimation seems to rely on the specific envelope kinematics, especially at the disk to envelope transition point. In this case group A lines like SO arise from both the disk and part of the envelope, while group B lines exclusively trace the disk. However, the disk-only model also gives a reasonable fit, and the fit can be potentially improved if different disk geometry or density profiles are to be used. In this case group A lines and group B lines are both tracing the disk, but group A lines like SO～$\rm 6_5-5_4$ 
has a more widespread distribution, likely due to their relatively low upper energy level (see \autoref{table:line_info}). It is difficult to distinguish the two different scenarios based on current observations. Higher resolution observations and further exploration of the envelope/disk properties, such as the geometry, density contrast between disk and envelope, are needed to clarify the issue. Given the systematic uncertainties involved in modeling the group A lines, we rely on the dynamical mass estimated for group B lines (compact disk tracers) in \autoref{sec:mass_estimation} for related discussions.


\startlongtable
\begin{deluxetable*}{ccccccccc}
\tabletypesize{\scriptsize}
\renewcommand{\arraystretch}{1.0}
\tablecaption{Estimated physical parameters of IRS3 from the kinematic modeling of SO $\rm 6_5-5_4$ \label{table:mcmc_env}}
\tablehead{
 \colhead{Type}  &\colhead{$m_*$} & \colhead{$R_{\rm disk}$} & \colhead{$R_{\rm in}$} & \colhead{$R_{\rm out}$}&\colhead{$ v_{\rm sys}$}  & \colhead{$\Delta x$} & \colhead{$m_{\rm env}/m_{\rm disk}$} &\colhead{$f_{\rm norm}$} \\
  & \colhead{($M_\odot$)} & \colhead{(au)} &\colhead{(au)} &  \colhead{(au)} & \colhead{(\kms)} & \colhead{(au)} &\colhead{} &\colhead{}
}
\startdata
   Disk       &	1.61$\pm$0.00&	269.04$\pm$1.57&	15.28$\pm$0.46& -&	9.96$\pm$0.01&	-8.39$\pm$0.54&-&	1.13$\pm$0.01\\
  Disk+Envelope(A)       &	1.24$\pm$0.04&	113.20$\pm$9.92&	10.76$\pm$2.78&	393.86$\pm$12.85&	9.67$\pm$0.15&	-11.79$\pm$1.81&	-&1.18$\pm$0.04\\
  Disk+Envelope(B)       &	1.44$\pm$0.09&	144.23$\pm$31.29&	14.98$\pm$1.79&	409.65$\pm$14.07&	9.65$\pm$0.19&	-9.08$\pm$1.48&	-&1.19$\pm$0.04\\
  Disk+Ulrich76 Envelope     &	1.43$\pm$0.10&	153.44$\pm$21.74&	12.57$\pm$1.85&	347.07$\pm$53.24&	9.56$\pm$0.20&	-8.48$\pm$1.25&0.51$\pm$0.02&	1.18$\pm$0.05\\
\enddata
\end{deluxetable*}

\section{Discussion}\label{sec:diss}

\begin{figure*}[ht!]
\epsscale{1.0}\plotone{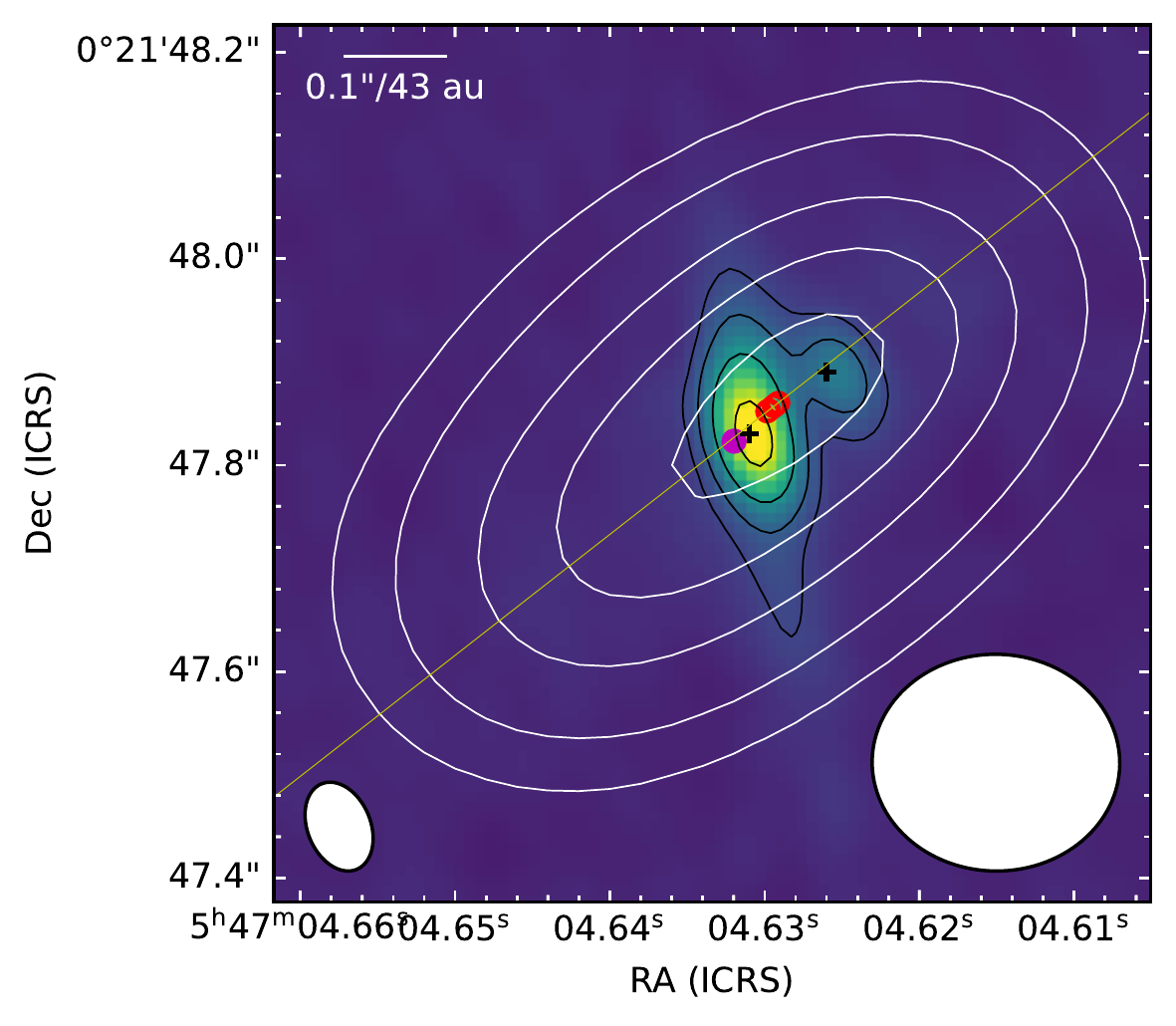}
\caption{A zoom-in view of IRS3. The colorscale and black contours illustrate the VLA 9~mm continuum. The contours levels are (10, 20, 40, 80)$\times \sigma$ with $\sigma$ = 12~\mujypbm. The white contours indicate the 1.3~mm continuum and the contour levels are (100, 200, 400, 600, 800)$\times \sigma$ with $\sigma$ = 1.3~\mjypbm. The beam sizes of 9~mm and 1.3~mm data are shown in the lower and right corner, respectively. The geometric center of the circumbinary disk, determined as the intensity weighted center of the region between 100$\sigma$ and 400$\sigma$ contours in 1.3~mm, is shown in a magenta dot. The positions of IRS3A and IRS3B, measured from the 9~mm data is indicated by black crosses. The red circles represent the ``kinematic center'' from the kinematic modeling measured for different lines.}
\label{fig:IRS3_zoom}
\end{figure*}

\subsection{Implication of the protostellar mass}
\label{sec:mass}

Despite of the fact that NGC~2071~IR is known to be an intermediate mass star formation region based on luminosity considerations, there is no consensus with regard to the protostar masses of the brighest sources, i.e., IRS1 and IRS3. \citet{Snell86} suggests that a single B2 star is required to generate sufficient ionizing flux to account for the observed radio emission of IRS1, but the majority of radio emission may arise from thermal jets rather than free-free emission of a photoionized HII region. \citet{Carrasco12} argues that IRS3 should also host an intermediate mass star by modelling its SED and spatial intensity profile at 3~mm with an irradiated accretion disk model. More constraints are derived from observations of water masers with VLA and VLBA by several investigators \citep{Torrelles98,Seth02,Trinidad09}. Based on the spatial-kinematic distribution of the water masers \citet{Trinidad09} estimated a central mass of 5$\pm$3~\msun{} and 1.2$\pm$0.4~\msun{} for IRS1 and IRS3, respectively. Nevertheless, the estimation is derived with only a small number of masers (5 for IRS1, 6 for IRS3) and relies on assumptions about the disk inclination and radii. Our ALMA molecular line observations provide a unique chance to clarify the dynamical masses of IRS1 and IRS3. 

\subsubsection{IRS3}
For IRS3 our kinematic modeling gives a central mass estimation of 1.43 -- 1.52~\msun{} from fits to different molecules. The variation is probably reflecting systematic differences of molecules in the spatial distribution within the disk, due to variation of abundances and excitation conditions. A more detailed modeling involving physically realistic disk properties and astrochemical evolution is required to better understand the variations of different molecules but is beyond the scope of this paper. Here we take the range 1.43 -- 1.52~\msun{} as a reasonable estimation for possible central masses of IRS3.  This estimation is further consolidated by our SED fitting in \autoref{sec:sed}, which favors a central mass around 2~\msun. Note that our VLA 9~mm observations have revealed the multiplicity in IRS3. In principle, if the molecular lines are tracing Keplerian orbiting motions around the two components (IRS3A, IRS3B) then the estimated dynamical mass should be treated as the sum of two. 

Our kinematic modeling also has the ability to constrain the mass ratio of IRS3A/IRS3B. In \autoref{sec:kine} we attempted the modeling of IRS3 with the position offset as a free parameter to allow for a precise measurement of the ``kinematic center'', which corresponds to the binary barycenter as the disk kinematics are regulated by the center of mass. In the fixed-$i$ case, the best-fit offsets range from $-$13.60 to $-$19.54~au for different lines, indicating the kinematic center is about 13.60 -- 19.54~au to the NW direction compared with the reference point, as illustrated in \autoref{fig:IRS3_zoom}. Therefore the kinematic center is located roughly in the middle of IRS3A/B and one can further estimate a IRS3A/IRS3B mass ratio of 1.3--2.5 by comparing their relative distances from the kinematic center. 

It is also interesting to note that IRS3A appears to coincide with the ``geometric center'' of the disk. Here the ``geometric center'' can be defined as the center of symmetry of elliptical contours in 1.3~mm/0.87~mm continuum at relatively low levels, where the emission is not obviously skewed to the NW direction. In \autoref{fig:IRS3_zoom} we marked the position of the intensity weighted center for outer disk (defined by the region between 100$\sigma$ - 400$\sigma$ isophotal contours in 1.3~mm). If the disk is in a steady state with its material distribution regulated by the central mass, one would expect its geometric center close to the barycenter of the binary system. However, in our case there seems to be a non negligible deviation between the two.


However, we note that our spatial resolution is only 0\farcs{23} ($\sim$100~au) in band 6 so it is challenging to infer the position of the kinematic center (or the relative position between ALMA/VLA detections) with a few au precision. One potential issue is the limitation in absolute positional accuracy, which hinders precise determination of the precise relative location between VLA/9~mm and ALMA/1.3~mm detections. The theoretical accuracy limit by signal to noise ratio can be estimated by $\delta p$ = $\theta_{\rm beam}/(S/N)/0.9$ \citep[e.g.,][]{Cortes20}, which is around 0.3~mas for a FWHM beam size $\theta_{\rm beam}$ = 0\farcs{23} and  S/N = 890 for IRS3 in band 6. However, the atmospheric phase fluctuations limit the $\delta p$ to about 0.05$\theta_{\rm beam}$ \citep{Cortes20}, which is around 0\farcs{02}, or 5~au. Secondly, the determination of the kinematic center relies on assumption of symmetrically distributed line emission on both sides of the disk. Nevertheless, we have observed asymmetric intensity distribution in high resolution 0.87~mm continuum, which is heavily skewed to the NW direction. If the asymmetry arises from local enhanced temperature or density (likely due to the existence of IRS3B), then the molecular line emission could be stronger toward the NW side as well, and thus leading to a deviation of the derived kinematic center to the NW direction. There are also some hints for asymmetry in line intensity distribution in some lines (see e.g, \autoref{fig:pv_IRS3}). Follow up observations in future are required to clarify it.

\subsubsection{IRS1}
The mass of IRS1 is less well constrained mainly due to a lack of knowledge of the inclination angle. In the case of a fixed inclination of 60\arcdeg, the kinematic modeling gives a central mass in range of 4.40 -- 5.46~\msun{} with the scattering from different molecules. Systematically larger (smaller) masses can be obtained if smaller (larger) inclination angles are to be assumed.
The free-$i$ cases favor a modestly large inclination of 56\arcdeg -- 70\arcdeg.

One can also argue against a very small inclination of i $\lesssim$ 30\arcdeg (close to face-on configuration) for IRS1 based on the observed morphology/kinematics of outflows driven by IRS1. The \htwo{} 1 - 0 S(1) map reveals a strong outflow associated with IRS1 roughly in the E-W direction, with the Eastern lobe extending as far as 30\arcsec ($\sim$0.06~pc) (outflow II in \citet{Eisloffel00}, see also \citet{Walther19}). In our CO 2--1 observations the outflow associated with IRS1 mainly appears as a single blueshifted lobe located to the Southwest of the source, in contrast with the face-on configuration for which one would expect more spatially overlapping blueshifted and redshifted line emission. Therefore the IRS1 disk should have a moderate or higher inclination, although we cannot be more certain about the precise range.

The SED analysis in \autoref{sec:sed} provides a good constraint on the upper limit of the central mass of IRS1. All the best five fit models of IRS1 have a stellar mass $m_* \lesssim$ 4~\msun. A larger stellar mass will typically result in larger fluxes in mid-/far-infrared, and bolometric luminosities compared with the observation. The best fit model with larger stellar mass (i.e., $m_*$ = 8~\msun) has a $\chi^2 \gtrsim$ 10, significantly worse compared with the minimum $\chi^2$ ($\sim$ 5), and hence is highly unlikely. Nevertheless, the ZT model grid is rather sparse for smaller $m_*$ and it is difficult to derive a more precise upper limit from the SED fitting. Combining the information from both kinematics and SED, we speculate the stellar mass of IRS1 is about 3 -- 5~\msun. This puts constraints on the inclination angle of IRS1, i.e., i $\gtrsim$ 60\arcdeg{} according to \autoref{table:mcmc}. A caveat is that in the SED analysis we have implicitly assumed that IRS1 is dominated by a single protostar while multiple components could exist inside IRS1. 

\subsection{Accretion rate}

In \autoref{sec:sed} our SED modeling returns an estimation of the accretion rate onto the protostar $\dot{m_*}$, which spans from 1.5 $\times$ 10$^{-5}$ to 1.9 $\times$ 10$^{-4}$~\msunyr{} for both IRS1 and IRS3. Much of the large uncertainties arise from emission being blended at wavelength $>$70~$\mu$m and hence the bolometric luminosity for either source is not well constrained. The bolometric luminosity $L_{\rm bol}$ has been estimated to be 478~$L_{\odot}$ in an aperture encompassing both IRS1 and IRS3 in \citet{Furlan16}. If we simply estimate the ratio of the $L_{\rm bol,IRS1}$/$L_{\rm bol,IRS3}$ based on the {\it SOFIA} 37.1~$\mu$m photometry, as done in \autoref{sec:cont}, then the $L_{\rm bol}$ of IRS1 and IRS3 are 368~$L_{\odot}$ and 85~$L_{\odot}$, respectively. Vast majority of the $L_{\rm bol}$ in IRS3 should be attributed to accretion luminosity since the central object has a mass smaller than 2~\msun{} based on our kinematic modeling \citep[see e.g.,][]{Palla93}. Assuming $L_{\rm acc} \approx L_{\rm bol}$, then the accretion rate can be estimated from the equation

\begin{equation}
    L_{acc} = \frac{Gm_{*}{\dot{m_*}}}{R_{ps}},
\end{equation}
where G is the gravitation constant, $m_{*}$ is the protostar mass, $\dot{m}_{*}$ is the mass accretion rate from the disk to the protostar, and $R_{ps}$ is the protostellar radius. Here we adopt a stellar radius of 5~$R_{\odot}$ based on the protostellar structure models in \citet{Palla93}. Thus the mass accretion rate is likely between 8.8 $\times 10^{-6}$ to 9.4 $\times 10^{-6}$~\msunyr{} for a protostellar mass in range of 1.43 -- 1.52~\msun. In this calculation we have assumed the accretion luminosity of IRS3 is dominated by a single protostar, whereas in \autoref{sec:mass} we have shown that the secondary component, IRS3B, may have a comparable mass as IRS3A, although the uncertainty in the estimated mass ratio is large. 
 From the above equation, the estimated $\dot{m}_{*}$ will not change if both components have a similar $\dot{m}_{*}$ and a $R_{\rm ps}$ of 5~$R_{\odot}$, regardless of their mass ratios. In this case the estimated $\dot{m}_{*}$, 8.8$\times 10^{-6}$  -- 9.4$\times 10^{-6}$~\msunyr, should be understood as accretion occurring onto both objects. It is difficult, though, to precisely determine the relative strength of $\dot{m}_{*}$, and accordingly $L_{\rm acc}$, for IRS3A/B since it may depend on their masses, as well as their locations in the disk. And their radii are probably smaller than 5~$R_{\odot}$ given their smaller masses.

This estimation of $\dot{m}_{*}$ is smaller (a factor $\gtrsim$ 2) compared with that derived in our SED modeling. However, in our SED modeling the returned $L_{\rm bol}$ for all best five models (100 -- 300~$L_{\odot}$) are greater than the assumed $L_{\rm bol}$ (85~$L_{\odot}$) here, possibly because for $\lambda>$70~$\mu$m only upper limits of the photometry measurements are given. For the solution with $L_{\rm bol}$ = 100~$L_{\odot}$, the ZT model gives a $\dot{m_*}$ of 1.5 $\times$ 10$^{-5}$~\msunyr, i.e., consistent within a factor of 2 with the derived value from simplified calculations. This remaining discrepancy mainly arises from a different treatment of the accretion luminosity in the ZT SED model, in which half of the accretion energy, i.e., ${Gm_{*}{\dot{m}_{*}}}/{2R_{ps}}$ is released when the accretion flow reaches the stellar surface, while the other half is partly radiated from the disk and partly converted to the kinetic energy of the disk wind. So up to half of the accretion luminosity may be converted to the kinematic energy and cannot be observed in radiation, thus a larger accretion rate is needed in ZT model to account for the accretion luminosities.

It is difficult to derive the accretion rate for IRS1 in the same way, due to the less well constrained dynamical mass and unknown multiplicity. If IRS1 contains two low mass protostars with mass $\lesssim$~2~\msun, then one can derive an accretion rate of 1.5 $\times 10^{-5}$~\msunyr{} following similar arguments as we did for IRS3, i.e., assuming $L_{\rm acc} \approx L_{\rm bol}$ and $R_{ps}$ = 5~$R_{\odot}$. However, IRS1 could consist of one or multiple protostars with higher masses, which may have larger stellar radii and a larger fraction of the observed luminosity could be dominated by stellar radiation instead of accretion. Indeed, in three out of the best five SED models for IRS1 in \autoref{sec:sed} with $m_*$ = 4~\msun, vast majority of the total luminosity is contributed by the protostar itself. But still we expect active accretion is currently occurring onto IRS1 based on the detection of high velocity CO outflows close to IRS1.

\subsection{Jets and outflows in NGC~2071~IR}
\label{sec:discussion_outflow}

Jets and outflows provide a fossil record of the mass-loss histories of associated YSOs.
The NGC~2071~IR region is characterized by widespread molecular hydrogen emission as revealed by the \htwo{} 1-0 S(1) line, and considerable efforts have been made to identify individual outflows and assigning individual protostars to them \citep{Eisloffel00,Walther19}. IRS3 appears to be the driving source for the largest NE-SW outflow that extends $\sim$3\arcmin{} far on both sides (outflow IA, IB following the designation in \citet{Eisloffel00}). IRS1 is driving another outflow that is more E-W oriented (outflow IIA, IIB, with PA$\sim$70\arcdeg), although the western lobe (IIB) is much fainter in \htwo{} lines. Our CO 2-1 observations, for the first time, provide a high resolution view of the molecular outflows within the central 30\arcsec{} in the NGC~2071~IR region. We discuss IRS3 and IRS1 separately in the following sections.

\subsubsection{IRS3}

The CO 2--1 data reveals a high velocity bipolar jet associated with IRS3, with a position angle in a range of 22 -- 32\arcdeg{} seen in different velocities. This further confirms the association of the large scale NE-SW \htwo{} outflow with IRS3. Interestingly, there is a clear misalignment between the jet/outflow seen in different tracers: the \htwo{} outflow IA/IB has an average position angle of $\sim$ 45\arcdeg, with its lateral extents covering a wide range around 10\arcdeg -- 60\arcdeg. The molecular jet shows a position angle in a range around 22 -- 32\arcdeg. The radio jet in our 9~mm map has a position angle $\sim$ 15\arcdeg (which is consistent with measurements in literature, see \citet{Torrelles98, Seth02, Trinidad09,Carrasco12}). Furthermore, the disk has a position angle of $\sim$ 130\arcdeg, approximately perpendicular to the high velocity CO jet (but not the radio jet). The most likely mechanism to account for these observational phenomena is a precessing jet wiggling over a range of jet position angles, and thus the various orientations seen in different tracers represent the interaction of jet and environment material at different phases. Jet precession has long been known in both low mass and high mass YSOs \citep[see reviews in][]{Frank14,Lee20}, and large axis changes of up to $\sim$45\arcdeg{} are also observed in a few sources \citep[e.g.,][]{Cunningham09}.

The current radio jet seen in centimeter continuum indicates the most recent ejection event from the protostar. It is also suggested in \citet{Carrasco12} that the jet is precessing since they found variations in the jet orientation (a few degrees) of the IRS3 jet over a few years (1995--1999). However, their observations have relatively low resolutions ($\sim$ 0.3\arcsec{} in 3.6~cm), and hence the measurements may suffer from more uncertainties due to the imperfect Gaussian fitting, or contaminated by the unresolved binary component. Our high resolution 9~mm data gives a more unambiguous measurement of the jet position angle, i.e, $\sim$ 15\arcdeg, which is close to the value measured for 1998/1999 epochs in \citet{Carrasco12}. There is no smoking gun evidence for the changes of radio jet orientation in a few year timescale, and further observations with similar spatial resolutions are needed to confirm/clarify it and better constrain the precession period. 

In some velocities the CO outflow appears as misaligned segments, and these features are roughly symmetrically distributed in blueshifted and redshifted lobes (e.g., at $|v - v_{\rm sys}|$ = 30~\kms, see \autoref{fig:outflow_chan}). Such point-symmetric (i.e., S-shaped) wiggles are expected for precession of the accretion disks because of tidal interactions in noncoplanar binary systems \citep[see, e.g.,][]{Raga93,Terquem99}. Similar misaligned segments and S-shape wiggle are also tentatively detected in the \htwo{} emission \citep[see figures in][]{Eisloffel00,Walther19}, but the spatial distribution of \htwo{} is more complex and may involve other physical processes including deflection or blocking of part of an outflow. At lower velocities we detected a wide angle component of the CO outflow. Interestingly, the edges of this component line up well with the spatial extent of the \htwo{} emission. This further consolidates a unified origin of the CO and \htwo{} emission and suggest that this wide outflow cavity is at least partly shaped by the jet precession.

\subsubsection{IRS1}

Combining the information in the literature and this work, IRS1 exhibits a similar complexity on the inferred ejection directions. \citet{Trinidad09} detected a few ejected condensations (IRS1E, 1W) from IRS1 along the East-West direction (P.A. $\sim$ 100\arcdeg) based on VLA 1.3~cm and 3.6~cm observations \citep[see also][]{Carrasco12}. In \htwo{} emission, IRS1 seems to drive a East-oriented outflow (P.A. $\sim$ 70\arcdeg). This outflow contains several shock knots and some interspread diffuse emission and are collectively called Outflow IIA in \citet{Eisloffel00}, but there could be contribution from other protostars like IRS2. We have also detected some blueshifted CO emission that is likely associated with Outflow IIA. The western lobe (IIB) is much fainter in \htwo. 
Note that given the disk orientation (P.A.$\sim$135\arcdeg) of IRS1, it it likely that the large scale outflow in the NE-SW direction seen in single dish CO observation \citep[e.g.,][]{Stojimirovic08}, is partly contributed by IRS1.
On the other hand, a smaller P.A. ($\sim$ 45\arcdeg) is required to account for 
the low velocity CO outflow, especially the blueshifted wide angle component to the southwest. There are some relatively high velocity CO knots distributed close to the E-W orientation as well. The situation is further complicated by the non-ideal disk appearance seen in high resolution 0.87~mm, 1.3~mm and 9~mm continuum. The relatively diffuse part of 0.87~mm emission appears consistent with a NE-SW oriented disk (P.A. $\sim$ 135\arcdeg) but it contains a bright inner part elongated at around P.A. $\sim$ 25\arcdeg. Extension along similar direction is also seen in 9~mm continuum, as well as a protuberance to the east. This East-orientated extension could be due to a E-W jet, as suggested in \citet{Trinidad09} based on ejected radio knots, whereas the extension with P.A. $\sim$ 25\arcdeg{} is less clear without auxiliary information.

In summary, the observed radio condensations or high velocity CO suggests a jet along or close to the E-W direction, while the disk kinematics and CO outflow agree with a NE-SW orientated ejection direction. The \htwo{} emission has a spatial content with a P.A. range approximately in between but may have contributions from other YSOs. Similar to the case of IRS3, one might consider a precessing jet to account for the observed ejection events from IRS1. \citet{Carrasco12} reported that the direction of IRS1 jet appear to be changing slightly over a few years based on the 3.6~cm continuum morphology. In particular, the jet slightly bends to the northwest around 0\farcs{5} to the west of the protostar.  \citet{Carrasco12} argued that it could be due to either the superposition of a binary jet or a single jet interacting with the ambient medium. The latter scenario is reminiscent of our 1.3~mm continuum map, where we have also observed a dust clump around 0\farcs{8} to the west of IRS1. This dense clump could be responsible for the bending feature in radio jet. Optionally, we note that a precessing jet can also naturally explain the variations in jet orientation and morphology.

The possibility of unresolved multiplicity cannot be ruled out with current data. \citet{Carrasco12} suggests that the extension in their 0.09\arcsec{} resolution 1.3~cm map can be interpreted as a marginally resolved close binary system with a separation $\sim$ 40~au. Our 9~mm map of IRS1 exhibits a similar protuberance to the east.
A secondary component could be responsible for the origin of precession. Alternatively, one could assign the observed ejections in different directions to different components in the binary system, e.g., one along E-W and the other along NE-SW. 
However, more aggressive weighting of the long baselines in the
imaging process does not present any definitive evidence of multiplicity at present.
In either case, IRS1 is an interesting target to follow up with higher resolution interferometer observations to resolve possible multiplicity and to investigate the accretion and jet ejection process associated with intermediate mass protostars.

\subsection{Possible interactions during cluster formation}

In the preceding sections we have demonstrated that both IRS1 and IRS3 show indication of interesting disk substructures, e.g., the IRS1 disk resembles a bar-spiral configuration while IRS3 is a circumbinary disk hosting a close binary system. In \autoref{fig:overview} we also see that some disks including IRS3 appear to connect with some filamentary diffuse emission. Some of these signatures could potentially be linked to the clustered environment in this intermediate mass star forming region. Dynamical interactions between young stars are common in the molecular cloud \citep{Bate18}, and could dramatically affect the structure and evolution of protostellar disks \citep[e.g.,][]{Pfalzner03,Winter18,Cuello20}. 

One can estimate the average time required for a particular star to undergo an encounter with another star passing within a pericenter distance, $d_{\rm min}$ as \citep{Davies03}

\begin{equation}
    \tau_{\rm enc} \simeq 33~{\rm Myr} \left(\frac{100~{\rm pc}^{-3}}{n} \right) 
    \left( \frac{v_\infty}{\rm 1~km~s^{-1}} \right) 
    \left( \frac{10^3~\rm AU}{d_{\rm min}} \right) 
    \left( \frac{M_\odot}{M_t} \right),
\end{equation}
where $n$ is the stellar number density in pc$^{-3}$, $v$ the mean relative velocity at infinity of the cluster stars and $M_t$ is the total mass of the stars involved in the encounter. Therefore, disks around more massive protostars are more susceptible to close encounters. For NGC~2071~IR, there are nine protostellar objects within $R_c \sim$0.03pc, yielding a number density of 8$\times10^4~{\rm pc}^{-3}$. For $d_{\rm min} \sim$ 1000~au and $v \sim$ 1~\kms (typical velocity dispersion in NGC~2071~IR \citep{VanKempen12}), the average encounter time is about 2$\times 10^4$~yr for a low mass protostar such as IRS3, with $M_t \sim$ 2\msun. This is comparable to the cluster crossing time $t_c$ = $R_c/v$ $\sim$ 3$\times 10^4$ yr, or the free-fall timescale $t_{\rm ff}$ = $\sqrt{3\pi/(32G\rho)}~\sim~3\times 10^4$ yr, where the density $\rho$ is estimated with a total mass of $\sim$30\msun{}(combining the gas mass of 21.7\msun{} in \citet{VanKempen12} and stellar masses) within a slightly larger radius of $\sim$0.05~pc. Thus it can not be ruled out that the properties of IRS1 or IRS3 disks may have been impacted and shaped by encounters with other young stars during the cluster formation process.

\section{Conclusion}\label{sec:con}

We have utilized ALMA/VLA data in conjunction with previous near- to far-infrared and single-dish submillimeter data to characterize the protostellar content of the intermediate mass star formation region, NGC~2071~IR, and in particular, the dominant sources IRS1 and IRS3. These observations allow for a detailed characterization of the properties of protostellar disks, jets/outflows and multiplicity associated with IRS1 and IRS3. The main findings are summarized as follows:


\begin{itemize}
\item[1.] 
IRS3 shows a clear disk appearance at 0.87~mm and 1.3~mm, which has a measured radius of 103~au and an inclination of $\sim$67\arcdeg. The 9~mm continuum observation further reveals a close binary system separated by $\sim$43~au. The more luminous component, IRS3A, is coincident with the geometric center of the disk, and drives a radio jet with a position angle around 15\arcdeg.
\item[2.] 
IRS1 is marginally resolved in 1.3~mm with a 0\farcs{24}$\times$0\farcs{21} beam. With a 0\farcs{13}$\times$0\farcs{10} resolution in 0.87~mm IRS1 appears to contain both a inner brighter component and a larger diffuse component, with approximately orthogonal orientations. IRS1 is marginally resolved and has a T-shape extension in 9~mm.
\item[3.] 
Both IRS1 and IRS3 exhibit clear velocity gradient across their protostellar disks in multiple spectral lines, indicating Keplerian rotation. Inspection of the PV diagrams suggests that the molecular lines can be divided into two groups: group A, including \ceighteeno, \htwoco, \so, show bright emission peak in the first and third quadrant extending to $\gtrsim$ 1\arcsec{} and may have contribution from both disk and envelope, while lines group B, including \methanol, \thirteenmethanol, \sotwo{} and other organic molecules appears as a continuous linear feature crossing the first and third quadrant without low velocity emission extending beyond 0\farcs{5} ($\sim$200~au) and are hence exclusively tracing the inner disk.

\item[4.]
We use the \citet{Zhang18} model to fit the SED of IRS1 and IRS3 from near-IR to millimeter wavelength. A reasonably good fit can be obtained with a stellar mass of $\sim$4~\msun{} for IRS1, and $\sim$2~\msun{} for IRS3.

\item[5.] 
We developed an analytic modeling and MCMC method to better constrain the dynamical masses of the central objects of protostellar disks. IRS3 is estimated to have a total dynamical mass of 1.43--1.52~\msun.
By comparing the relative separation of each binary component (IRS3A, IRS3B) from the kinematic center determined in the modeling, we estimated a IRS3A/IRS3B mass ratio of 1.3 -- 2.5.
The dynamical mass is less clear for IRS1 without reliable measurement of the inclination.
In the free-$i$ case, the kinematic modeling gives a range of inclination from 56\arcdeg{} to 70\arcdeg, and mass from 3.58 to 5.65~\msun. Combining the constraints from both SED and kinematic modeling, IRS1 should have central mass in the range 3 -- 5~\msun{} assuming it is dominated by a single protostar.

\item[6.]
We modeled the group B line, SO $\rm 6_5-5_4$, with different setups, including a pure Keplerian disk and a disk plus an inner envelope, where different envelope models are investigated. The disk + simple envelope (A) model, in which we adopt the radius of the centrifugal barrier as the transition point of the disk and envelope, appears to lead to an underestimation of the central stellar mass, and is thus not favored.

\item[7.]
Based on our CO 2--1 data, IRS3 drives a spectacular high velocity jet, as well as a low velocity wide angle outflow. IRS1 drives a single lobe bubble-like outflow, as well as a few high velocity clumps. For both IRS1 and IRS3, the inferred ejection directions from different tracers, including radio jet, water maser, molecular outflow and \htwo{} emission, are not always consistent. For IRS1 the disagreement can be as large as $\sim$50\arcdeg. IRS3 is better explained with a single precessing jet. Similar mechanism may be working in IRS1 as well but unresolved multiplicity cannot be ruled out.
\end{itemize}

\vspace{5mm}
\facilities{Atacama Large Millimiter/submillimeter Array (ALMA)}
\software{CASA \citep{McMullin07}, APLpy \citep{Robitaille12}, Astropy \citep{Astro13}}

\acknowledgements
Support for this work was provided by the NSF through the Grote Reber 
Fellowship Program (to Y.C.) administered by Associated Universities,
Inc./National Radio Astronomy Observatory. J. J. T. acknowledges
funding from NSF grant AST-1814762.
Y.-L. Y acknowledges
the support from the Virginia Initiative of Cosmic
Origins (VICO) Postdoctoral Fellowship.
M.L.R.H acknowledges support from the Michigan Society of Fellows.
Z.-Y. L. is supported in part by NASA 80NSSC20K0533 and NSF AST-1815784.
G.A. and M.O. acknowledge support from the Spanish MINECO/AEI through the AYA2017-84390-C2 grant (co-funded by FEDER), MCIN/AEI/10.13039/501100011033 through the PID2020-114461GB-I00  and  from the State Agency for Research of the Spanish MCIU through the “Center of Excellence Severo Ochoa” award for the Instituto de Astrof{\'i}sica de Andaluc{\'i}a (SEV-2017-0709). 
MO also acknowledges financial support from the Consejería de 
Transformación Económica, Industria, Conocimiento y Universidades of the 
Junta de Andalucía and the European Regional Development Fund from the 
European Union through the grant P20-00880.
This paper makes use of the following ALMA data:
ADS/JAO.ALMA\#2015.1.00041.S and ADS/JAO.ALMA \#2018.1.01038.S.
ALMA is a partnership of ESO (representing its member
states), NSF (USA) and NINS (Japan), together with NRC (Canada), MOST and ASIAA
(Taiwan), and KASI (Republic of Korea), in cooperation with the Republic of Chile.
The Joint ALMA Observatory is operated by ESO, AUI/NRAO and NAOJ. The
National Radio Astronomy Observatory is a facility of the National Science
Foundation operated under cooperative agreement by Associated Universities, Inc.

\appendix
\counterwithin{figure}{section}
\counterwithin{table}{section}

\section{SED in the millimeter/centimeter regime}
\label{sec:app_sed}
\begin{figure}[ht!]
\epsscale{1.0}\plotone{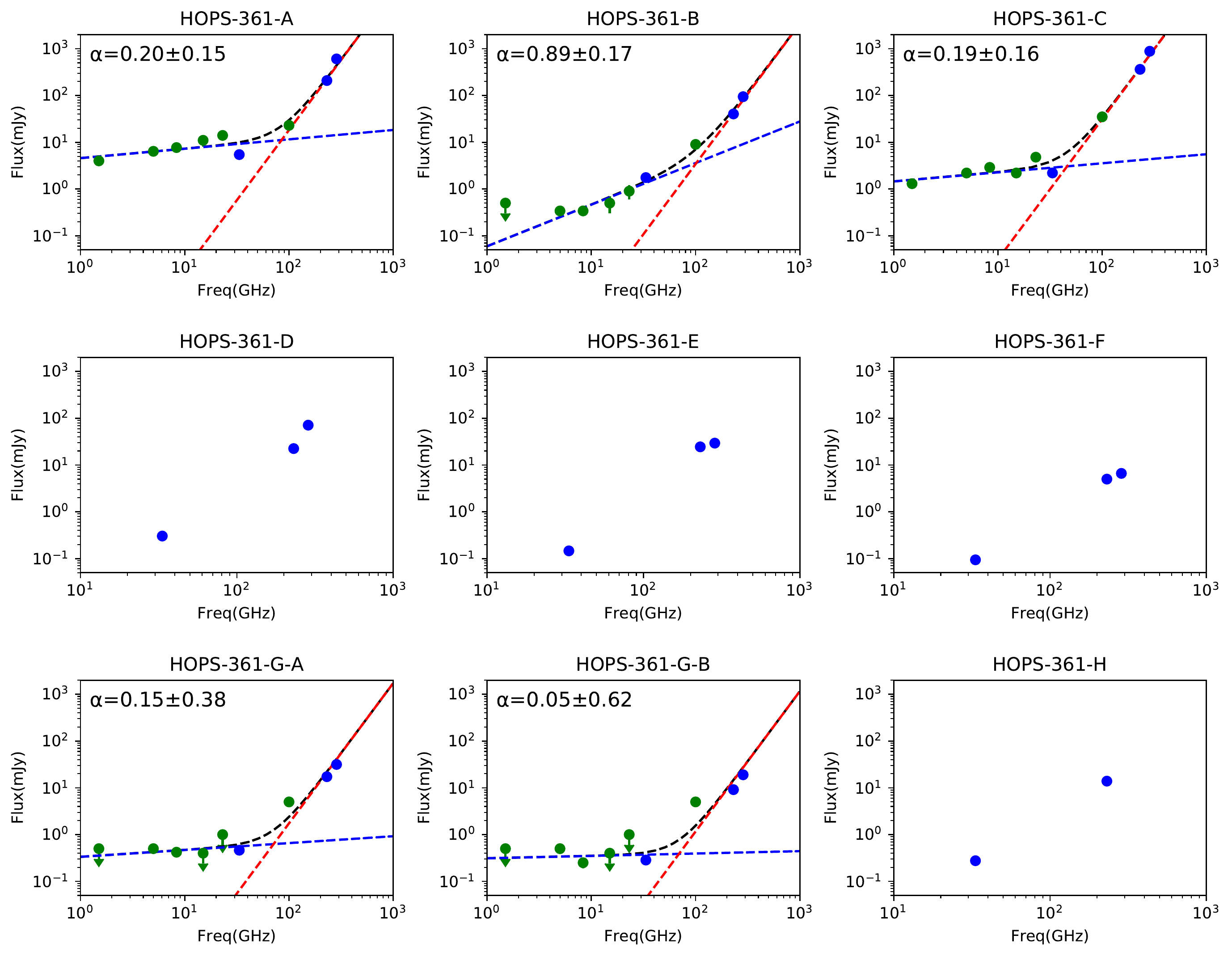}
\caption{SED of protostellar sources from 0.87~mm to 20~cm. The data points have been fitted as the sum of two powerlaws, one of them with a fixed slope of $+$3 (for dust). The green data points are collected from \citet[][ and see reference therein]{Carrasco12} and the blue data points are from this work.}
\label{fig:sed_long}
\end{figure}

In \autoref{fig:sed_long}, we show SED of protostars from 0.87~mm to 20~cm, with data in this work and also flux measurements in \citet{Carrasco12}. For sources with more than three data points we attempt a SED fitting with two powerlaw components, with one of them having a fixed slope of $+$3 (for thermal dust emission). Most sources show typical radio spectra found in YSOs, which are consistent with free–free emission at cm wavelengths plus a thermal dust contribution at millimeter wavelengths.

\section{{\it Xclass} modelling of the spectra in the continuum spectral window}
\label{sec:app_lines}
\begin{figure}[ht!]
\epsscale{1.1}\plotone{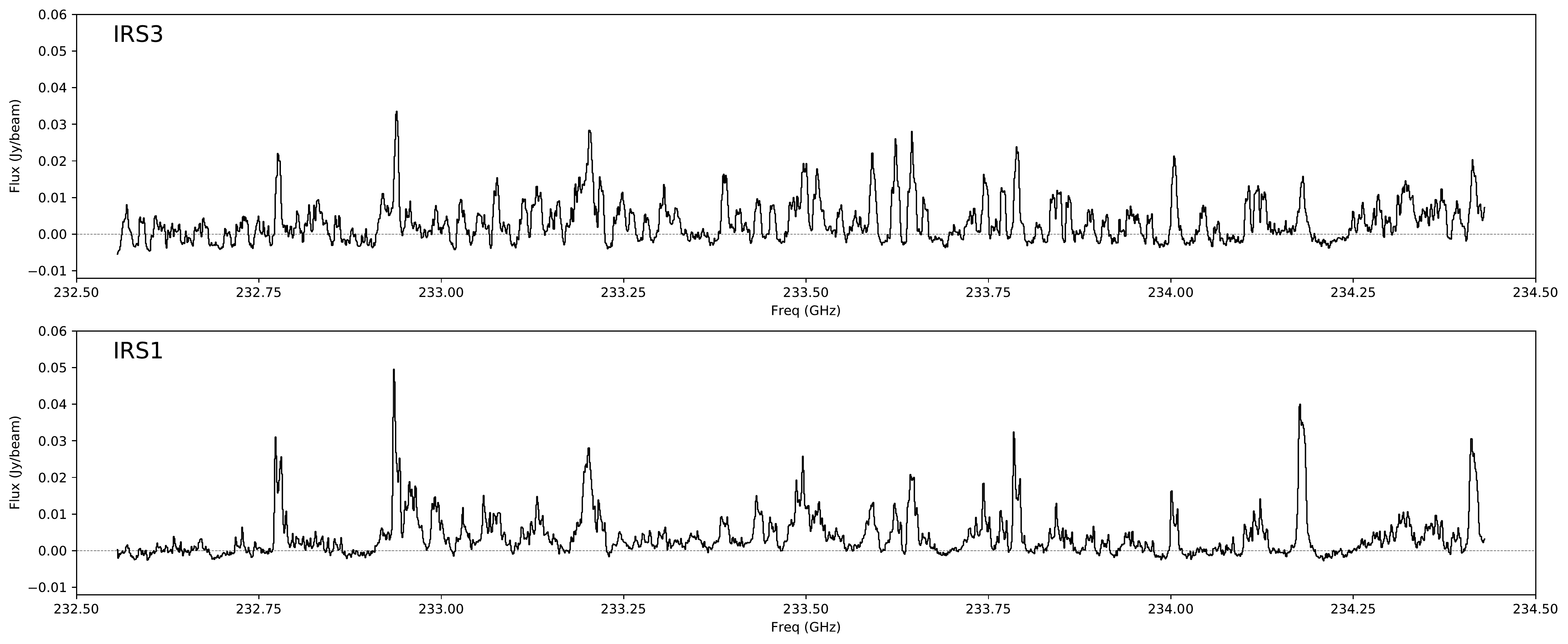}
\caption{Spectra from the {\it ALMA} band 6 continuum spectral window, extracted at the location of IRS1 and IRS3.}
\label{fig:spec}
\end{figure}

We modeled the emission of complex organic molecules to confirm the identifications and derive their column densities. To capture the emission from the vicinity of the protostars, we used a circular aperture with a radius of the maximum $R_{\rm disk}$ in \autoref{table:mcmc}. 
An overview of the spectra is shown in \autoref{fig:spec}. IRS1 exhibits clear doubled-peaked line profiles for most transitions due to disk kinematics and likely also dust opacity. Similarly double-peaked (or flat top) profiles are present in IRS3 as well (e.g., lines around 234.12~GHz) but there are also many lines appearing as single-peaked in the averaged spectrum.  


We used \textsc{xclass} \citep{Moller17},  which performs local thermodynamic equilibrium (LTE) radiative transfer calculations using the molecular data from CDMS and JPL, to identify molecular species and estimate their column densities.  \autoref{tbl:molcat} lists the molecular catalogs used in the modeling.  We set up a source model described as a thin disk that has four parameters, source size, excitation temperature ($T_\text{ex}$), column density ($N$), and the full line width at half maximum (FWHM).  We fix the source size as 0\farcs{5} and measured the FWHM from representative emission in each source.  For IRS1, we use a FWHM of 14.4\,\kms\ fitted from the SO$_2$ emission at 234187 MHz; for IRS3, we use a FWHM of 7.6\,\kms\ fitted from the \methanol\ emission at 232946 MHz.  We assume no continuum emission for the modeling.  To identify molecular species, we look for transitions with Einstein-A $> 10^{-6}$ s$^{-1}$ and upper energy $<$ 500 K.  Once a species is tentatively identified, we modeled its spectra with fiducial parameters, $T_\text{ex} = $100 K and a column density that produces emission similar to observations, to confirm that all detectable transitions appears in the observations. If a species has its modeled spectra consistent with the observations but all transitions are blended with the emission of other species, we consider it a tentative identification.  

\begin{figure}[htbp!]
    \centering
    \includegraphics[width=\textwidth]{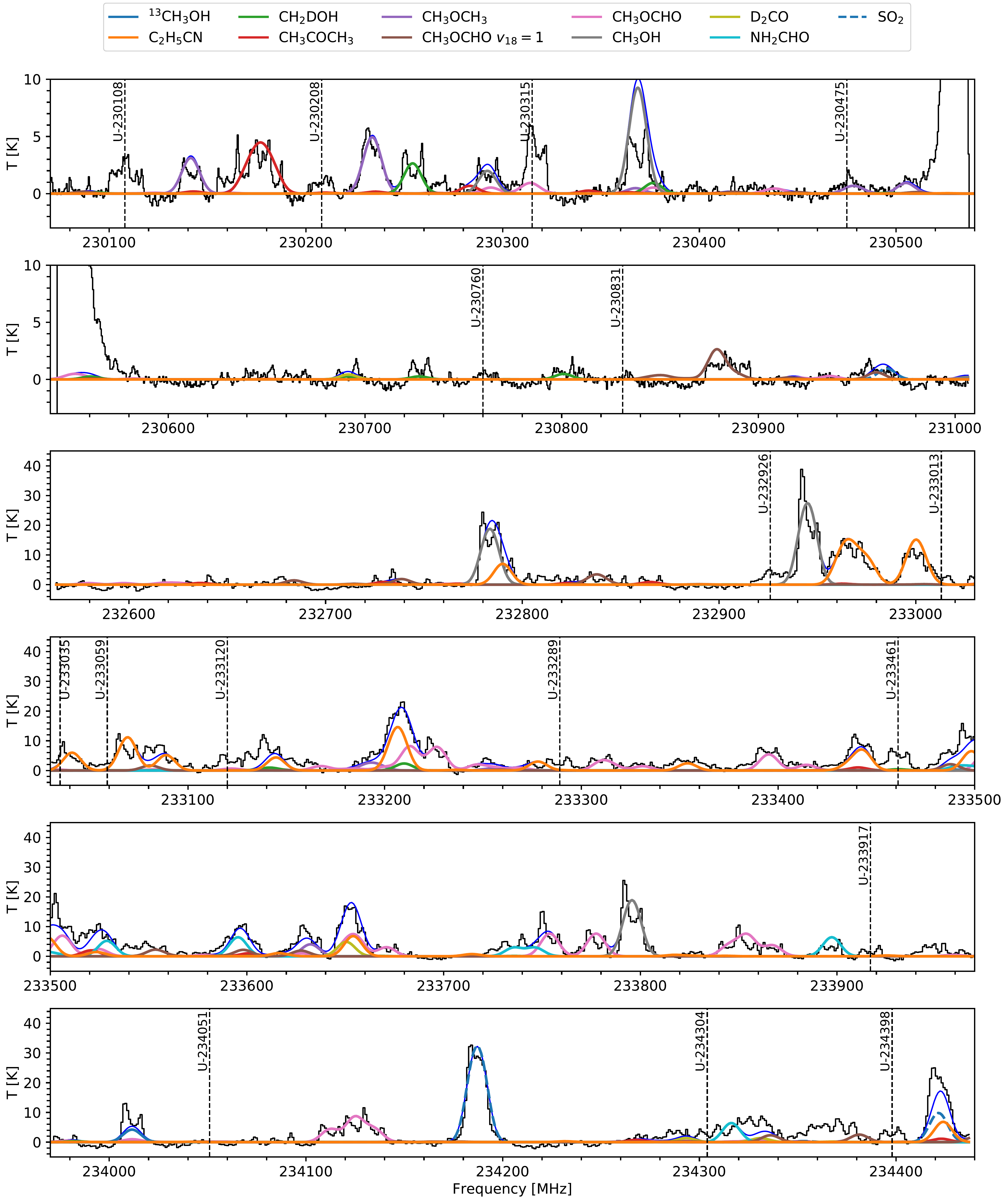}
    \caption{The spectra of IRS1 (black) and the synthetic spectra of COMs (color lines).  The total synthetic spectra is shown in thin blue lines. The vertical dashed lines indicate unidentified lines with their frequencies annotated.}
    \label{fig:irs1}
\end{figure}

\begin{figure}[htbp!]
    \centering
    \includegraphics[width=\textwidth]{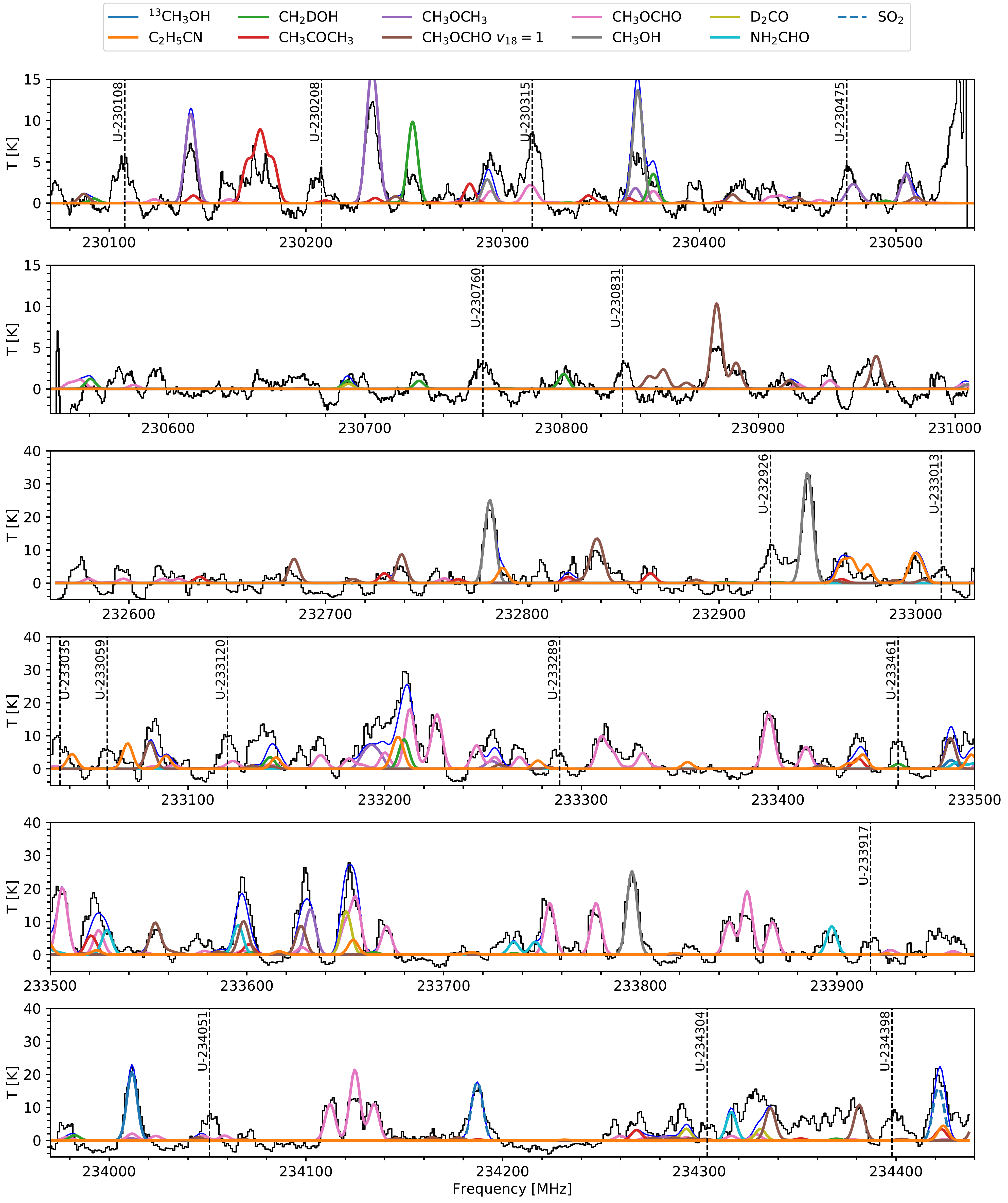}
    \caption{The spectra of IRS3 and the synthetic spectra of COMs. The legends are similar to that in Figure\,\ref{fig:irs1}}
    \label{fig:irs3}
\end{figure}

Then, we fitted all identified and tentatively identified species simultaneously to constrain their $T_\text{ex}$ and $N$. To reduce degeneracy in modeling, we first fitted the spectra of \methanol\ and \methylformate\ using their isolated emission prior to the global optimization that includes all species.  Then, we set the $T_\text{ex}$ of \dmethanol\ and \tmethanol\ to the fitted $T_\text{ex}$ of \methanol\ because there are only a few lines of these isotopologues detected in the observations. Finally, we ran a global optimization with all identified species without re-optimizing the models of \methanol\ and \methylformate. \autoref{tbl:irs_fitting} lists the best-fitted parameters for the identified species, while \autoref{fig:irs1} and \autoref{fig:irs3} show the synthetic spectra compared with the observations.

\begin{figure*}[ht!]
\epsscale{0.8}\plotone{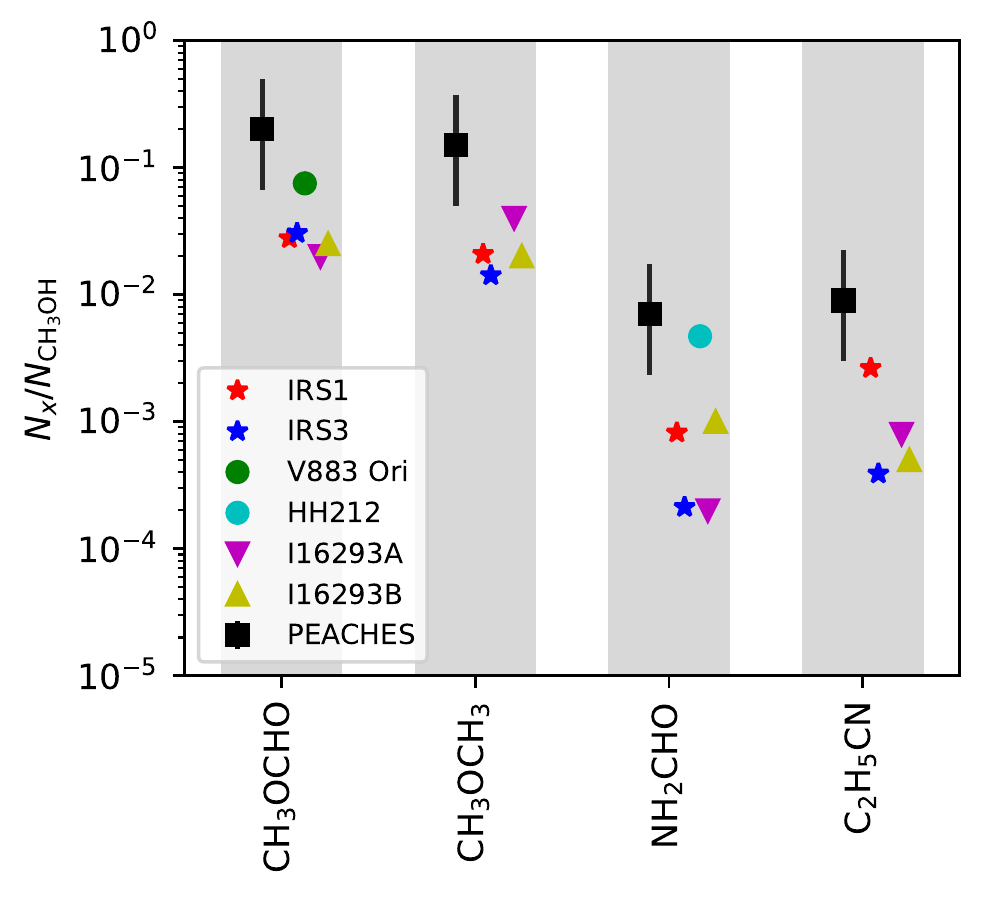}
\caption{
The ratios of the COMs detected in IRS1 and IRS3 compared to other studies of protostars.
The black lines indicate the abundance ratios of COMs toward the PEACHES sample \citep{Yang21}, which includes 50 embedded (Class 0/I) protostars in the Perseus molecular cloud. IRAS 16293−2422 is a prototype hot corino system. The column densities of COMs toward IRAS~16293-2422~B are taken from \citet{Jorgensen16} for \methanol, \citet{Jorgensen18} for CH$_3$OCH$_3$, CH$_3$OCHO, NH$_2$CHO and \citet{Calcutt18} for C$_2$H$_5$CN. The column densities of COMs toward IRAS~16293-2422~A are taken from \citet{Manigand20} for \methanol, C$_2$H$_5$OH, CH$_3$OCH$_3$, CH$_3$OCHO, NH$_2$CHO, and \citet{Calcutt18} for C$_2$H$_5$CN. V883~Ori and HH212 are two protostellar systems with spatially resolved COM detections associated with their disks. Their COM column densities are taken from \citet{Lee19} and \citet{Lee17}, respectively.\label{fig:com}
}
\end{figure*}

Most COMs in IRS1 and IRS3 are spatially resolved and show kinematics consistent with Keplerian rotation. Hence our measurements provide valuable constraints on the abundance of COMs that are exclusively associated with the disk. In \autoref{fig:com} we compare the COM abundances to other protostellar systems. Only a couple of tracers, for which measurements of other sources are also available in the literature, are shown here. The abundance is normalized by the column density of \methanol. Since the column density estimation of \methanol{} could be affected by optical depth we assume the true \methanol{} column density is the column density of \thirteenmethanol{} multiplied by the elemental abundance of $^{12}C/^{13}C$ = 60 \citep{Langer93}. The abundance ratios of IRS1 and IRS3 are generally lower, by about 0.5 -- 1.5 magnitude, compared to the PEACHES survey towards protostars in the Perseus molecular cloud \citep{Yang21}, or the protostellar disks HH212 and V883~Ori \citep{Lee17,Lee19}. Compared with the archetype hot corinos like IRAS~16293-2422~B, the abundance ratios of IRS1 or IRS3 are generally consistent. Note that the interpretation of such comparisons could be hindered by limited number of molecules and large systematic uncertainties in deriving the column densities in different works. Interestingly, while IRS1 and IRS3 exhibit similar abundance ratios for $\rm CH_3OCHO$ and $\rm CH_3OCH_3$, the ratio of IRS1 is 4 -- 7 times higher than IRS3 in $\rm NH_2CHO$ and $\rm C_2H_5CN$. Such differentiation may point to different behaviors in N-/O-bearing COMs, and reflect different initial conditions/chemical evolutionary stages in the two systems. 


\begin{table*}[htbp!]
  \caption{Molecular Catalogs}
  \label{tbl:molcat}
  \centering
  \begin{tabular}{cp{2in}cp{2in}}
  Molecule & References & Molecule & References \\
  \toprule
  \methanol & \citet{Xu08} & 
  \dmethanol & \citet{Pearson12} \\
  \tmethanol & \citet{Plummer84,Oesterling99,Carvajal07,Maeda08,Ilyushin09} & 
  D$_2$CO & \citet{Fabricant77,Dangoisse78,chardon1974spectre,Tucker73,Baskakov88,Lohilahti04} \\
  \methylformate\ ($v=0, 1$) & \citet{Ilyushin09} &
  \dimethylether & \citet{Endres09} \\
  \ethylcyanide & \citet{Pearson94,Brauer09} &
  \acetaldehyde & \citet{Kleiner96} \\
  \formamide & \cite{Blanco06,Kryvda09} & 
  SO$_2$ & \citet{Patel79,Helminger85,Lovas85,Alekseev96} \\
  \bottomrule
  \end{tabular}
\end{table*}

\begin{table}[htbp!]
    \centering
    \caption{Fitted Column Densities and $T_\text{ex}$}
    \begin{tabular}{c|ccc|ccc}
        \toprule
        Molecule & \multicolumn{3}{c}{IRS 1} & \multicolumn{3}{c}{IRS 3} \\
                 & $T_\text{ex}$ (K) & $N$ (cm$^{-2}$) & tentative & $T_\text{ex}$ (K) & $N$ (cm$^{-2}$) & tentative \\
        \midrule
        CH$_{3}$OH & 248 & 2.4\ee{18} & & 280 & 2.1\ee{18} & \\ 
        $^{13}$CH$_{3}$OH & 248$^{a}$ & 1.2\ee{17} & & 280$^{a}$ & 4.7\ee{17} & \\ 
        CH$_{2}$DOH & 248$^{a}$ & 8.2\ee{16} & & 280$^{a}$ & 2.1\ee{17} & \\ 
        CH$_{3}$OCHO & 168 & 2.0\ee{17} & & 450 & 8.6\ee{17} & \\ 
        SO$_{2}$ & 201 & 3.1\ee{17} & & 86 & 4.5\ee{17} & \\ 
        NH$_{2}$CHO & 359 & 5.9\ee{15} & & 499 & 6.0\ee{15} & \\ 
        D$_2$CO & 406 & 1.6\ee{16} & x & 182 & 1.0\ee{16} & x \\ 
        CH$_{3}$COCH$_{3}$ & 68 & 1.4\ee{16} & & 51 & 3.1\ee{16} & \\ 
        CH$_{3}$OCH$_{3}$ & 113 & 1.5\ee{17} & & 110 & 4.0\ee{17} & \\ 
        C$_{2}$H$_{5}$CN & 304 & 1.9\ee{16} & & 403 & 1.1\ee{16} & \\ 
        \bottomrule
        \multicolumn{5}{p{4in}}{$^{a}$The $T_\text{ex}$ is fixed.} \\
    \end{tabular}
    \label{tbl:irs_fitting}
\end{table}

\bibliography{refer}
\end{document}